\documentclass[onecolumn,prd,showpacs,preprintnumbers,amsmath,amssymb,floatfix]{revtex4}

\usepackage[pdftex]{graphicx}

\newcommand{\pom}{\tt I\! P}
\newcommand{\beq}{\begin{equation}}
\newcommand{\eeq}{\end{equation}}

\def \pom {{I\!\!P}}

\begin{document}

\title{Photon and Pomeron -- induced production of Dijets in $pp$, $pA$ and $AA$ collisions}

\author{E. Basso$^1$, V. P. Goncalves$^{2}$, A. K. Kohara$^{3,4}$ and  M. S. Rangel$^{3}$ }
\affiliation{ $^{1}$ Faculdade de Ci\^encias Exatas e Tecnologias,
Universidade Federal da Grande Dourados (UFGD),\\
Caixa Postal 364, Dourados, CEP 79804-970, MS, Brazil.}
\affiliation{$^{2}$ Instituto de F\'{\i}sica e Matem\'atica,  Universidade
Federal de Pelotas (UFPel), \\
Caixa Postal 354, CEP 96010-900, Pelotas, RS, Brazil}
\affiliation{$^{3}$ Instituto de F\'isica, Universidade Federal do Rio de Janeiro (UFRJ), 
Caixa Postal 68528, CEP 21941-972, Rio de Janeiro, RJ, Brazil\\}
\affiliation{$^{4}$ {Centro Brasileiro de Pesquisas F\'isicas  (CBPF), 
CEP 22290-180, Rio de Janeiro, RJ, Brazil}\\}

\begin{abstract}
In this paper we present a detailed comparison of  the dijet  production by photon -- photon, photon -- {pomeron} and {pomeron} -- {pomeron}  interactions in {${pp}$, ${pA}$ and ${\rm AA}$} collisions at the LHC energy. The transverse momentum, pseudo -- rapidity and angular dependencies of the cross sections are calculated at LHC energy using the Forward Physics Monte Carlo (FPMC), which allows to obtain realistic predictions for the dijet production with  two leading intact hadrons.  We obtain that $\gamma \pom$ channel is dominant at forward rapidities in {${pp}$} collisions and in the full kinematical range in the nuclear collisions of heavy nuclei. Our results indicate that the analysis of dijet production at the LHC can be useful to test the Resolved Pomeron model as well as to constrain the magnitude of the absorption effects.

\end{abstract}


\maketitle

\section{Introduction}

The  experimental results from Tevatron, RHIC and LHC for exclusive processes, characterized by a low hadronic multiplicity, intact hadrons and rapidity gaps in final state, has demonstrated that the study of these processes is feasible and that the data can be used to improve our understanding of the strong interactions theory as well constrain possible scenarios for the beyond Standard Model physics (For a recent review see, e.g. Ref. \cite{forward}).  In particular, it is expected that the forthcoming data can be used to discriminate between different approaches for the {pomeron}, which is a long-standing puzzle in the Particle Physics \cite{pomeron}. This object, with the vacuum quantum numbers, was introduced phenomenologically in the Regge theory as a simple moving pole in the complex angular momentum plane, to describe the high-energy behaviour of the total and elastic cross-sections of the hadronic reactions \cite{collins}. Due to its zero color charge, the {pomeron} is associated with diffractive events, characterized by the presence of large rapidity gaps in the hadronic final state. 

\begin{figure}[t]
\begin{center}
\begin{tabular}{cccc}
\scalebox{0.45}{\includegraphics{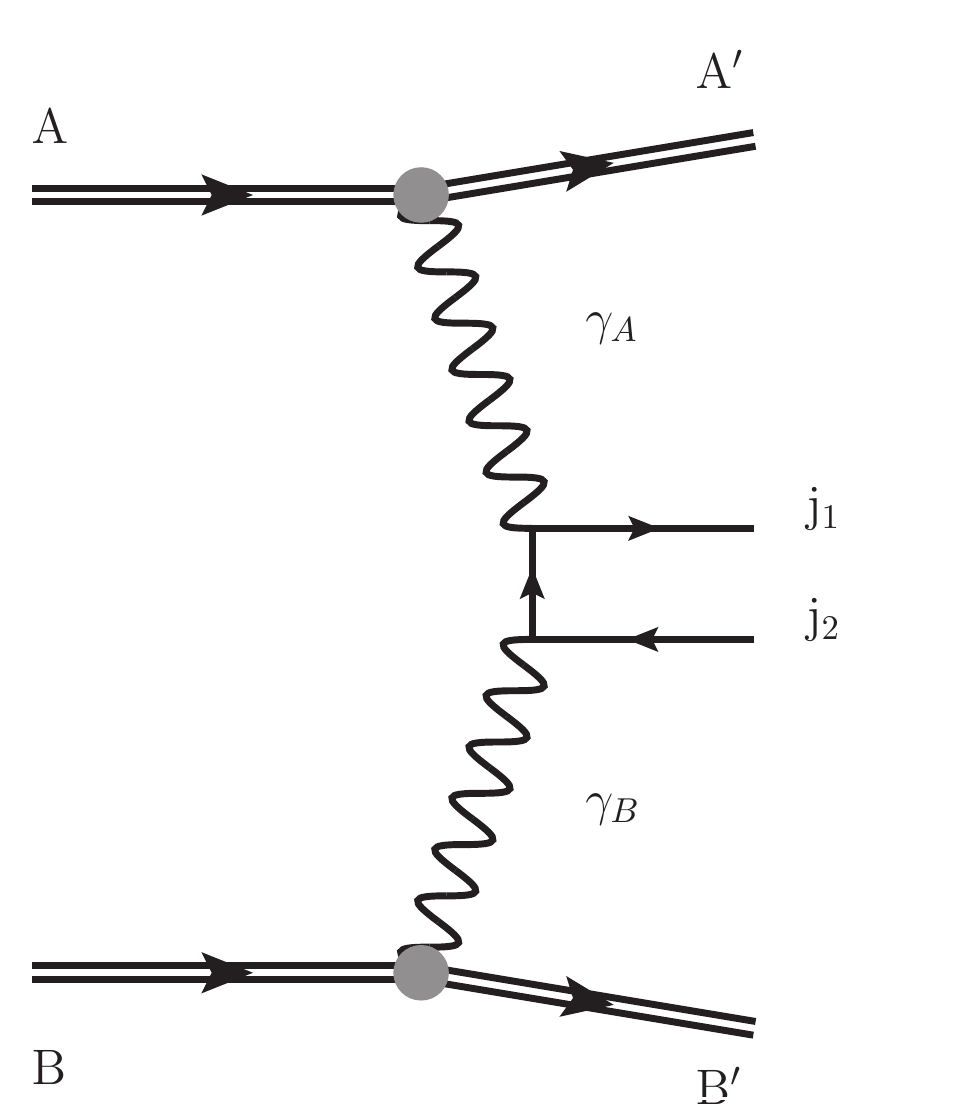}}&\scalebox{0.45}{\includegraphics{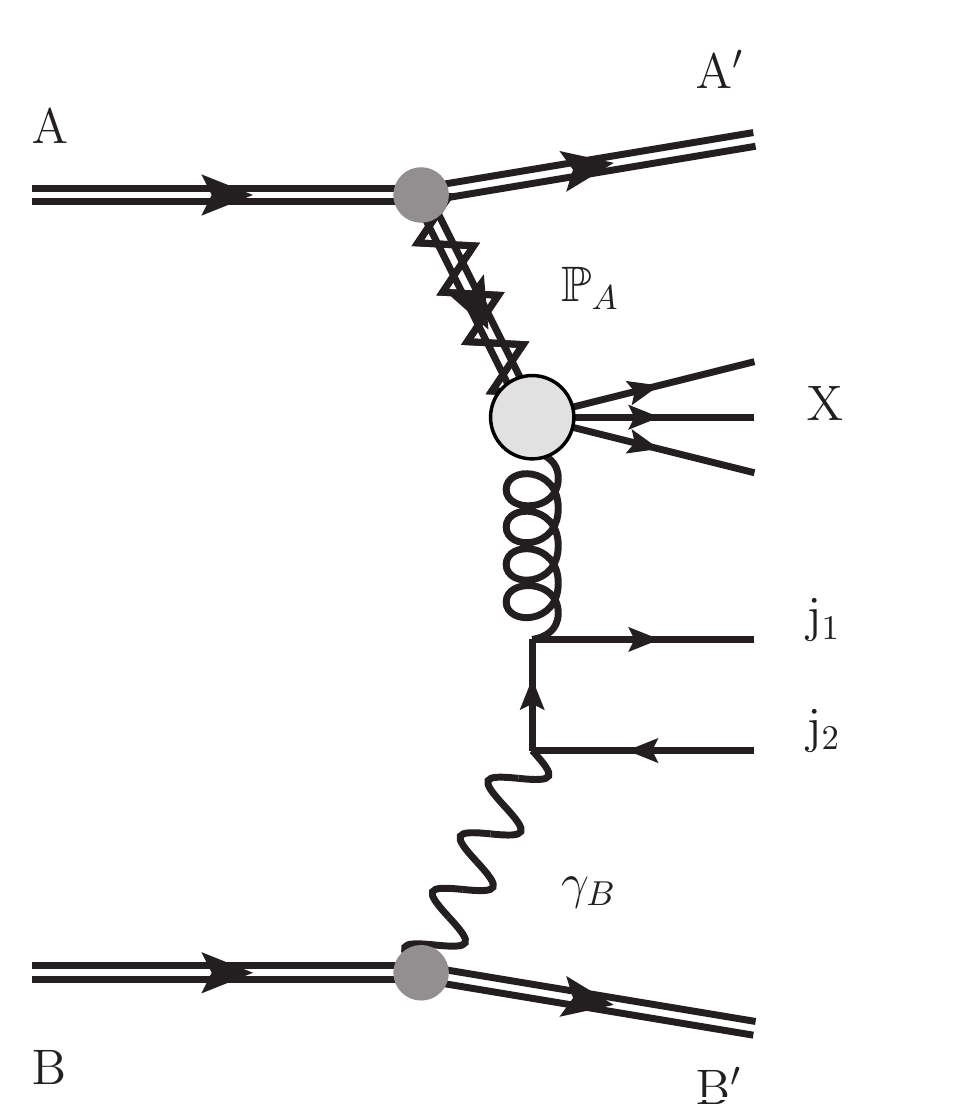}} &\scalebox{0.45}{\includegraphics{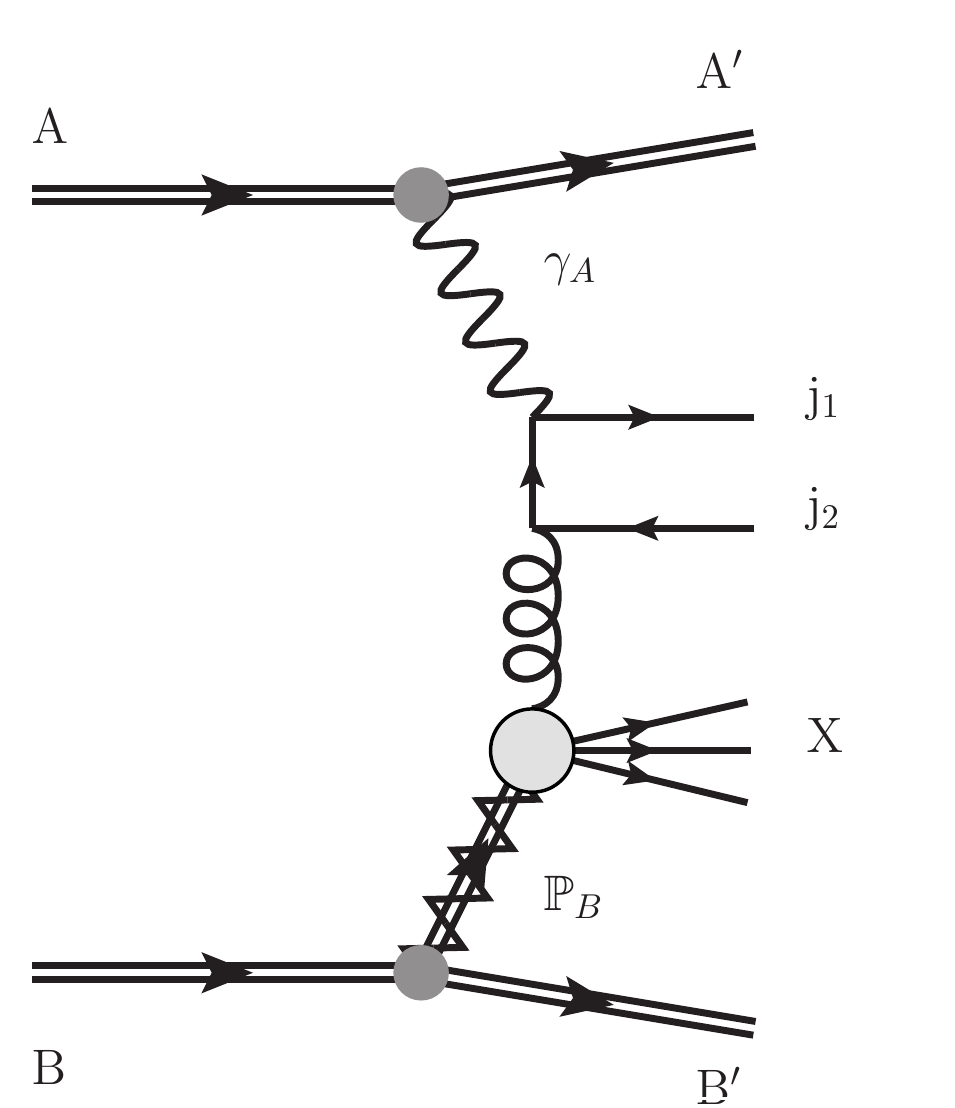}} & \scalebox{0.45}{\includegraphics{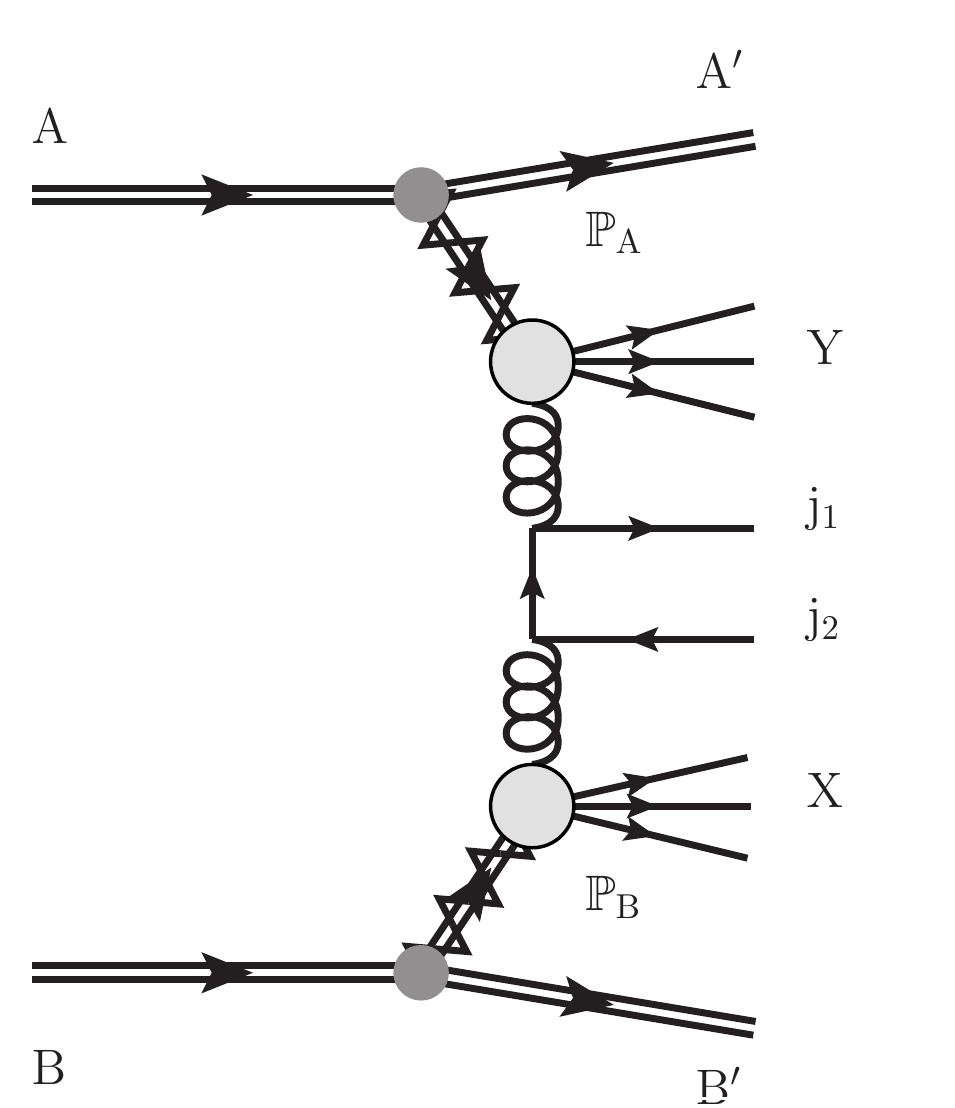}} \\
(a) & (b) & (c) & (d)
\end{tabular}
\caption{Dijet production by (a) photon -- photon, (b) pomeron -- photon, (c) photon -- pomeron and (d) pomeron -- pomeron  interactions in hadronic collisions.}
\label{Fig:dia}
\end{center}
\end{figure}

One good testing ground for diffractive physics and for the nature of the {pomeron} ($\pom$), is the dijet production in hadronic collisions. This process provides important tests of perturbative QCD and is one of the most important backgrounds to new physics processes. These aspects have  motivated the development  of an extensive phenomenology for this process in the last years 
\cite{prospects,covolan,cudell,marquet1,roman,cristiano,marquet2,kohara_marquet,
torb,marquet3,antonirecent}. In particular,  dijet production by photon - {pomeron} interactions in ultraperipheral \,{${pp/pA/AA}$} collisions, characterized by two intact hadrons and two rapidity gaps in the final state,  has been recently investigated in Ref. \cite{vadim}, considering the Resolved Pomeron model, in which the \,{pomeron} is assumed to have a partonic structure, as proposed by Ingelman and Schlein  \cite{IS} many years ago. They have obtained large values for the cross sections as a function of various variables. The \,{promising}   results presented in Ref. \cite{vadim} motivate a more detailed analysis of the dijet production, taking into account the contribution of other {processes} that are characterized by the same topology. In what follows, we will estimate the dijet production in photon -- photon and \,{pomeron -- pomeron} interactions present in \,{${pp/pA/AA}$} collisions and compare the predictions with those for the dijet production in photon -- \,{pomeron} interactions. These different processes are represented in Fig. \ref{Fig:dia}, and as emphasized before, they are characterized by two hadron intacts in the final state as well as two rapidity gaps. One importance difference between the dijet production by $\gamma \gamma$ interactions and the other processes, is that in \,{pomeron} -- induced  processes, the Resolved Pomeron model predicts the \,{existence} of  particles accompanying the dijet, with the associated rapidity gaps becoming, in general, smaller than in the $\gamma \gamma$ case. Additionally, the photon and \,{pomeron} -- induced processes are expected to generate emerging hadrons with different transverse momentum distributions, with those associated to \,{pomeron} -- induced having larger transverse momentum. Consequently, in principle, it is possible to introduce a selection criteria  to separate these different contributions for the dijet production. Although these distinct processes have been studied separately by several groups in the last years, the calculations have been performed considering different approximations and assumptions, which makes  difficult the direct comparison between its predictions. Our goal in this paper is to estimate these processes considering the same set of assumptions for the \,{pomeron} and for the \,{photon} flux and  obtain realistic predictions for the dijet production in photon and \,{pomeron} -- induced interactions including experimental cuts in the  calculations. In order to do that, we will use the   Forward Physics Monte Carlo (FPMC),  proposed some years ago \cite{fpmc} to treat  \,{pomeron -- pomeron} and photon -- photon interactions in hadronic collisions and recently improved to also include photon -- \,{pomeron} interactions in \,{pp} collisions \cite{nosbottom}. Here we generalize this Monte Carlo to treat $\gamma \gamma$, $\gamma \pom$ and $\pom \pom$ interactions in \,{pA} and \,{AA} collisions. As a consequence, it is possible to  estimate the contribution of the different processes presented in Fig. \ref{Fig:dia} in a common framework. In this paper we will  perform a comprehensive analysis of the transverse momentum and pseudo -- rapidity distributions for the different processes.

The content of this paper is organized as follows. In the next section we present a brief review of the formalism for the dijet production in photon and \,{pomeron} -- induced interactions in hadronic collisions. In Section \ref{results} we present our predictions for the pseudo -- rapidity and transverse momentum distributions  for the dijet production in  \,{pp/pA/AA} collisions at LHC energies, considering the contributions associated to $\gamma \gamma$, $\gamma \pom$ and  $\pom \pom$ interactions. Finally, in Section \ref{conc} we summarize our main conclusions.


\section{Dijet production in photon and pomeron -- induced interactions}
\label{sec:dijet}

At high energies,  a ultra relativistic charged hadron (proton or nuclei)
 give rise to strong electromagnetic fields, such that the photon stemming from the electromagnetic field of one of the two colliding hadrons can interact with one photon of the other hadron (photon - photon process) or can interact directly with the other hadron (photon - hadron process) \cite{upc,epa}. In these processes the total cross section  can be factorized in terms of the equivalent flux of photons into the hadron projectiles and the photon-photon or photon-target production cross section.  In particular, the dijet production by $\gamma \gamma$ \,{interactions} at high energies in  hadronic collisions, represented in Fig. \ref{Fig:dia} (a), can be described at leading order by the following expression
\begin{eqnarray}
\sigma(h_{\rm A} h_{\rm B} \rightarrow h_{\rm A} \otimes j_1 j_2 \otimes h_{\rm B}) = \int dx_{\rm A} \int dx_{\rm B} \, \gamma_{\rm A}(x_{\rm A},\mu^2) \cdot \gamma_{\rm B}(x_{\rm B},\mu^2) \cdot \hat{\sigma}(\gamma \gamma \rightarrow j_1 j_2) \,\,,
\label{fotfot}
\end{eqnarray}
where $\gamma_i(x_i,\mu^2)$ is the equivalent photon distribution of the hadron $i$, with  $x_i$ being the fraction of the hadron energy carried by the photon and $\mu$ has to be identified with a hard scale of the  process. Moreover, $\otimes$ represents the presence of a rapidity gap in the final state and 
$\hat{\sigma}$ is the partonic cross section for the  $\gamma_{\rm A} \gamma_{\rm B} \rightarrow j_1 j_2$ subprocess. On the other hand, the cross section for the dijet production in photon -- pomeron  interactions, represented in Figs. \ref{Fig:dia} (b) and (c),  
is given by
\begin{eqnarray}
\sigma(h_{\rm A} h_{\rm B} \rightarrow h_{\rm A} \otimes j_1 j_2 X \otimes h_{\rm B} ) = \int dx_{\rm A} \int dx_{\rm B} \, [g^D_{\rm A}(x_{\rm A},\mu^2) \cdot \gamma_{\rm B}(x_{\rm B},\mu^2) + \gamma_{\rm A}(x_{\rm A},\mu^2) \cdot g^D_{\rm B}(x_{\rm B},\mu^2)] \cdot \hat{\sigma}(\gamma g \rightarrow j_1 j_2) \,\,, 
\label{pompho}
\end{eqnarray}
where $g^D_i (x_i,\mu^2)$ is the diffractive gluon distribution of the hadron $i$ with a momentum fraction $x_i$ and we take into account that both incident hadrons can be a source of photons and pomerons. Finally, the cross section for the dijet production in double diffractive processes, represented in Fig. \ref{Fig:dia} (d), can be expressed by
\begin{eqnarray}
\sigma(h_{\rm A} h_{\rm B} \rightarrow h_{\rm A} \otimes X j_1 j_2 Y \otimes h_{\rm B}) = \int dx_{\rm A} \int dx_{\rm B} \, g^D_{\rm A}(x_{\rm A},\mu^2) \cdot g^D_{\rm B}(x_{\rm B},\mu^2) \cdot \hat{\sigma}(g g \rightarrow j_1 j_2) \,\,,
\label{pompom}
\end{eqnarray}
 where, for simplicity, we assumed that the dominant subprocess is the $g g \rightarrow j_1 j_2$ interaction, which is a good approximation at high energies. However, in the numerical calculations, the contribution associated to the $q \bar{q} \rightarrow 
j_1 j_2$ subprocess also have been included.

The basic ingredients in the analysis of these photon and \,{pomeron} -- induced processes are the  equivalent photon distribution of the incident hadrons $\gamma(x,\mu^2)$  
and its diffractive gluon distributions $g^D (x,\mu^2)$. As our goal is to calculate the cross sections for the processes presented in Fig. \ref{Fig:dia} considering \,{pp, pA and AA} collisions, we should to specify the associated models used in the proton and nuclear cases. Initially, lets present the models used for the photon distribution.
 The equivalent photon approximation of a charged  \,{point-like} fermion was formulated  many years ago by Fermi \cite{Fermi} and developed by Williams \cite{Williams} and Weizsacker \cite{Weizsacker}.  In contrast, the calculation of the photon distribution of the hadrons still is a subject of debate, due to the fact that they are not \,{point-like} particles. In this case it is necessary to distinguish between the  elastic and inelastic components.  The elastic component, \,{$\gamma_{\rm el}$}, can be estimated analysing the transition $h \rightarrow \gamma h$ taking into account the effects of the hadronic form factors, with the hadron remaining intact in the final state \cite{epa,kniehl}. In contrast, the inelastic contribution, \,{$\gamma_{\rm inel}$}, is associated to the transition $h \rightarrow \gamma X$, with $X \neq h$, and  can be estimated taking into account the partonic structure of the hadrons, which can be a source of photons. In what follows we will consider the contribution associated to elastic processes, where the incident hadron remains intact after the photon emission (For a related discussion about this subject see Refs. \cite{vicgus1,vicgus2}). For the proton case, a detailed derivation of the  elastic photon distribution  was presented in Ref. \cite{kniehl}.  Although an analytical expression for the elastic component is presented in Ref. \cite{kniehl}, it is common to found in the literature the study of photon - induced processes considering an approximated expression for the photon distribution of the proton proposed in Ref. \cite{dz}, which can be obtained from the full expression by disregarding the contribution of the magnetic dipole moment and the corresponding magnetic form factor. As demonstrated in Ref. \cite{vicwerdaniel} the difference between the full and the approximated expressions is smaller than 5\% at low-$x$. Consequently, in what follows we will use the expression proposed in Ref. \cite{dz}, where the elastic photon distribution of the proton is given by
 \begin{eqnarray}
\gamma^{\rm el}_{p}(x)&=&\frac{\alpha_{\rm em}}{\pi}\left(\frac{1-x+0.5x^{2}}{x}\right) \times  \left[\ln(\Omega)
-\frac{11}{6}+\frac{3}{\Omega}-\frac{3}{2\Omega^{2}}+\frac{1}{3\Omega^{3}}\right]\,\,,
\label{dz}
\end{eqnarray}
where $\Omega=1+(0.71 \mathrm{GeV}^{2})/Q_{\rm min}^{2}$ and 
 $Q^2_{\rm min} \approx (x m)^2/(1-x)$.
On the other hand, the equivalent photon flux of a nuclei is assumed to be given by \cite{upc}
\begin{eqnarray}
\gamma^{\rm el}_{\rm A}(x)&=& \frac{\alpha_{\rm em}\,Z^2}{\pi}\,\frac{1}{x} \left[2\,\bar{\eta}\,K_0\,(\bar{\eta})\, K_1\,(\bar{\eta})- \bar{\eta}^2\,{\cal{U}}(\bar{\eta}) \right]\,\,\,,
\label{fluxint}
\end{eqnarray}
where   $\bar{\eta}= x \frac{\sqrt{s}}{2} \,(R_{h_{\rm A}} + R_{h_{\rm B}})/\gamma_L$, $R_{h_i}$ is the hadron radius  and  
${\cal{U}}(\bar{\eta}) = K_1^2\,(\bar{\eta})-  K_0^2\,(\bar{\eta})$. One have that $\gamma^{\rm el}_{\rm A}(x)$ is enhanced by a factor $Z^2$ in comparison to the proton one.
 
Lets now discuss the modelling of the diffractive gluon distributions for the proton and nucleus. In order to describe the diffractive processes we will consider in what follows the Resolved Pomeron model \cite{IS}, which assumes that the diffractive parton distributions  can be expressed in terms of parton distributions in the \,{pomeron} and a Regge parametrization of the flux factor describing the \,{pomeron} emission by the hadron. The  parton distributions have evolution given by the DGLAP evolution equations and should be determined from events with a rapidity gap or a intact hadron. In the proton case, the diffractive gluon distribution, $g^D_{p} (x,\mu^2)$, is defined as a convolution of the \,{pomeron} flux emitted by the proton, $f^{p}_{\pom}(x_{\pom})$, and the gluon distribution in the \,{pomeron}, $g_{\pom}(\beta, \mu^2)$,  where $\beta$ is the momentum fraction carried by the partons inside the \,{pomeron}. 
The \,{pomeron} flux is given by
\begin{eqnarray}
f^{p}_{\pom}(x_{\pom})= \int_{t_{\rm min}}^{t_{\rm max}} dt \, f_{\pom/{p}}(x_{{\pom}}, t) = 
\int_{t_{\rm min}}^{t_{\rm max}} dt \, \frac{A_{\pom} \, e^{B_{\pom} t}}{x_{\pom}^{2\alpha_{\pom} (t)-1}}  \,\,,
\label{fluxpom:proton}
\end{eqnarray}
where $t_{\rm min}$, $t_{\rm max}$ are kinematic boundaries. The \,{pomeron} flux factor is motivated by Regge theory, where the \,{pomeron} trajectory is assumed to be linear, $\alpha_{\pom} (t)= \alpha_{\pom} (0) + \alpha_{\pom}^\prime t$, and the parameters $B_{\pom}$, $\alpha_{\pom}^\prime$ and their uncertainties are obtained from fits to H1 data  \cite{H1diff}. \,{ The slope of the pomeron flux is $B_{\pom}=5.5^{-2.0}_{+0.7}$ GeV$^{-2}$, the Regge trajectory of the pomeron is
$\alpha_{\mathbb P}(t)=\alpha_{\mathbb P}(0)+\alpha_{\mathbb P}'~t$ with $\alpha_{\mathbb P}(0)=1.111 \pm 0.007$ and $\alpha_{\mathbb P}'=0.06^{+0.19}_{-0.06}$ GeV$^{-2}$. The $t$ integration boundaries are $t_{\rm max}=-m_{p}^2x_{\pom}^2/(1\!-\!x_{\pom})$ ($m_{p}$ denotes the proton mass) and $t_{\rm min}=-1$ GeV$^2$.} Finally, the normalization factor $A_{\mathbb P}=1.7101$ is chosen such that $x_{\pom}\times\int_{t_{\rm{min}}}^{t_{\rm{max}}}dt~f_{\pom/{p}}(x_{\pom},t)=1$ at $x_{\pom} = 0.003$.
The diffractive gluon distribution of the proton is then given by
\begin{eqnarray}
{ g^D_{p}(x,\mu^2)}=\int dx_{\pom}~d\beta ~\delta (x-x_{\pom}\beta)~f^{p}_{\pom}(x_{\pom})~g_{\pom}(\beta, \mu^2)={ \int_x^1 \frac{dx_{\pom}}{x_{\pom}} f^{p}_{\pom}(x_{\pom}) ~g_{\pom}\left(\frac{x}{x_{\pom}}, \mu^2\right)} \,\,.
\label{difgluon:proton}
\end{eqnarray}
Similar definition can be established for the diffractive quark distributions.
In our analysis we use the diffractive gluon distribution obtained by the H1 Collaboration at DESY-HERA, denoted fit B in Ref. \cite{H1diff}. However, we checked that similar results are obtained using the fit A.
In order to specify the diffractive gluon distribution for a nucleus $g^D_{\rm A} (x,\mu^2)$, we will follow the approach proposed in Ref. \cite{vadim} (See also Ref. \cite{review_vadim}), which estimate $g^D_{\rm A}$ taking into account the nuclear effects associated to the nuclear coherence and the leading twist nuclear shadowing. The basic assumption is that  the  pomeron - nucleus coupling is proportional to the mass number \,{A} \cite{berndt}. As the associated pomeron flux depends on the square of this coupling, this model predicts that when the pomerons are coherently emitted by the nucleus,  $f_{\pom/{\rm A}}$ is proportional to ${\rm A}^2$.  Consequently, the nuclear diffractive gluon distribution can be expressed as follows (For details see Ref. \cite{vadim})
\begin{eqnarray}
{ g^D_{\rm A}(x,\mu^2)}= R_g \, {\rm A}^2 \,  { \int_x^1 \frac{dx_{\pom}}{x_{\pom}} \left[ \int dt \, f_{\pom/{p}}(x_{{\pom}}, t) \cdot F_{\rm A}^2(t) \right]  g_{\pom}\left(\frac{x}{x_{\pom}}, \mu^2\right)} \,\,,
\label{difgluon:nucleo}
\end{eqnarray}
where $R_g$ is the suppression factor associated to the nuclear shadowing and $F_{\rm A}(t)$ is the nuclear form factor. In what follows we will assume that $R_g  = 0.15$ as in Ref. \cite{vadim} and that $F_{\rm A}(t) \propto e^{R_{\rm A}^2 t/6}$, with $R_{\rm A}$ being the nuclear radius. 

One important open question in the treatment of photon  and \,{pomeron} -- induced is if the cross sections for the associated processes are not somewhat modified by 
soft interactions which lead to an extra production of particles that destroy the rapidity gaps in the final state \cite{bjorken}. As these effects have nonperturbative nature, they are difficult to treat and its magnitude is strongly model dependent (For recent reviews see Refs. \cite{durham,telaviv}). In the case of $\pom \pom$ interactions in \,{${pp / p\bar{p}}$} collisions, the experimental results obtained at TEVATRON \cite{tevatron} and LHC \cite{atlas_dijet,cms_dijet} have demonstrated that   one should take into account of these additional absorption effects that imply the violation of the QCD hard scattering factorization theorem for diffraction \cite{collinsfac}. In general, these   effects are parametrized in terms of a rapidity gap survival probability, $S^2$, which corresponds to the probability of the scattered proton not to dissociate due to the secondary interactions. Different approaches have been proposed to calculate these effects  giving distinct predictions (See, e.g. Ref. \cite{review_martin}). An usual approach in the literature is the calculation of an average probability $\langle |S|^2\rangle$ and after to multiply  the cross section by this value. As previous studies for the double diffractive production \cite{nosbottom,MMM1,marquet3,antoni,antoni2,cristiano,cristiano2} we also follow this simplified approach assuming $\langle |S|^2\rangle = 0.02$ for the dijet production by $\pom \pom$ interactions in \,{pp} collisions. It is important to emphasize that this choice is somewhat arbitrary, and mainly motivated by the possibility to compare our predictions with those obtained in other analysis. Recent studies from the CMS Collaboration \cite{cms_dijet} indicate that this factor can be larger  than this value by a factor $\approx 4$. The magnitude of $\langle |S|^2\rangle$ for $\pom \pom$ interactions in \,{pA} and \,{AA} collisions is still more uncertain \cite{berndt,acta_martin,miller,radion}.
In what follows we will consider the approach proposed in Ref. \cite{berndt} for coherent double exchange processes in nuclear collisions. The basic idea in this approach is to express the $\pom \pom$ cross section in the impact parameter space, which implies that the double pomeron exchange process becomes dependent on the magnitude of the geometrical overlap of the two nuclei during the collision. As a consequence, it is possible to take into account the centrality of the incident particles and estimate $\langle |S|^2\rangle$ by requiring that the colliding nuclei remain intact, which is equivalent to suppress  the interactions at small impact parameters ($b < R_{\rm A} + R_{\rm B}$). In order to obtain predictions for $\langle |S|^2\rangle$ in  \,{pA} and \,{AA} collisions at  LHC energies, we have updated and improved the model proposed in Ref. \cite{berndt} and obtained the values presented in Table \ref{tab:pomnuc}. In Appendix A we give a brief explanation of the model for $\langle |S|^2\rangle$ calculation. A detailed discussion of the model will be presented in a separated publication. One have that the predicted values for $\langle |S|^2\rangle$ are larger than those obtained in Ref. \cite{miller} using a Glauber approach and in Ref. \cite{radion} assuming that the nuclear suppression factor is given by $\langle |S|^2 \rangle_{{\rm A}_1{\rm A}_2} = \langle |S|^2 \rangle_{pp}/({\rm A}_1.{\rm A}_2)$. Consequently, our predictions for the dijet production by $\pom \pom$ interactions in \,{pA/AA} collisions may be considered an upper bound for the magnitude of the cross sections. 
In the case of $\gamma \gamma$ and $\gamma \pom$ interactions,  we will assume  $\langle |S|^2\rangle = 1$, motivated by results obtained e.g. in Refs. \cite{bruno,brunorecente}, which verified that the recent LHC data for the exclusive vector meson production in photon -- induced interactions can be described without the inclusion of a normalization factor associated to absorption effects. However, it is important to emphasize that the magnitude of the rapidity gap survival probability in $\gamma \pom$ still is an open question. For example, in Ref. \cite{Schafer}  the authors have estimated $\langle |S|^2\rangle$ for the exclusive photoproduction of $J/\Psi$ in  \,{pp/p\=p} collisions, obtaining that it is $ \sim 0.8 - 0.9$ and depends  on the  rapidity of the vector meson (See also Refs. \cite{Guzey,Martin,vadim}). Therefore, similarly to our $\pom \pom$ predictions, the results for the dijet production by $\gamma \gamma$ and $\gamma \pom$ interactions also may be considered an upper bound.

\begin{table}[t]
\begin{center}
 \vspace{0.5cm}
\begin{tabular}{|c|c|c|c|}
\hline
\hline
 & $^{16}{\rm O}$ & $^{40}{\rm Ca}$ & $^{208}{Pb}$ \\
\hline
\hline
pA & 0.0288  & 0.0185 & 0.0123   \\
\hline
AA & 0.00084 & 0.00019 & 0.000034  \\
  \hline
\hline
\end{tabular}
\caption{Gap survival probability $\langle |S|^2 \rangle$ for $\pom \pom$ interactions in pA and AA collisions at $\sqrt{s} = 5.02$ TeV.}
\label{tab:pomnuc}
\end{center}
\end{table}

\begin{figure}[t]
\begin{center}
\begin{tabular}{ccc}
\includegraphics[scale=0.3]{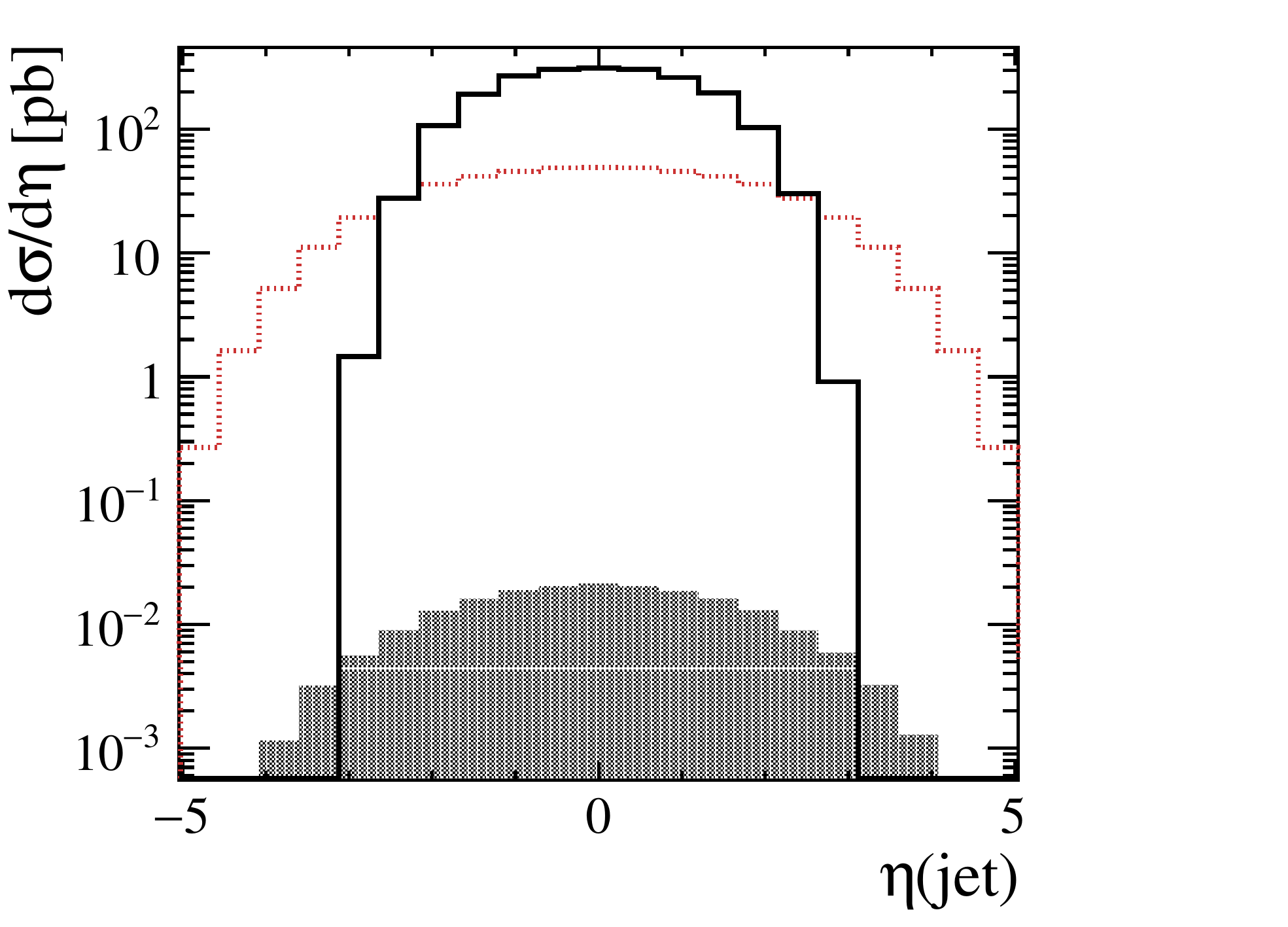}
&  \includegraphics[scale=0.3]{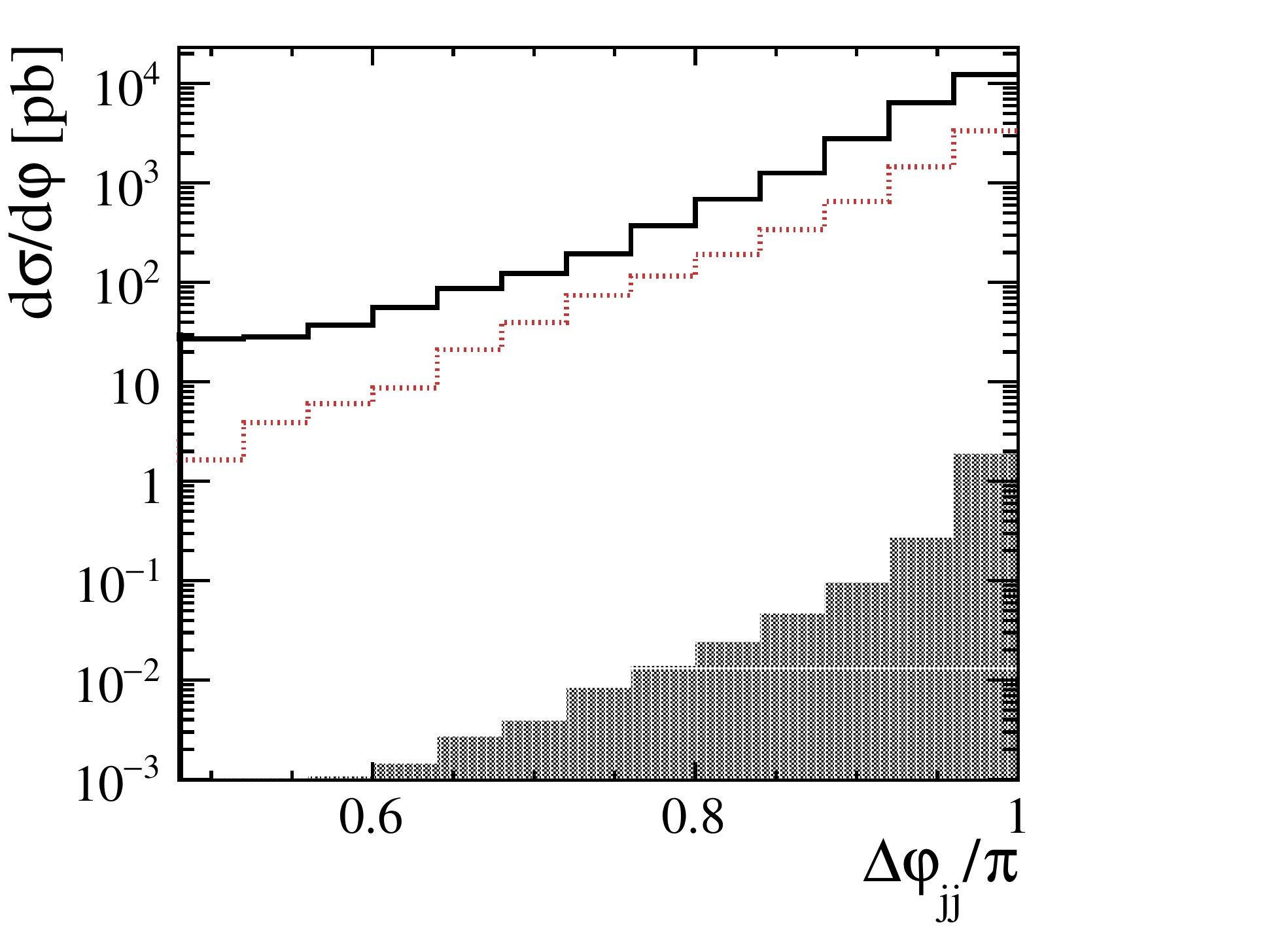} & \includegraphics[scale=0.3]{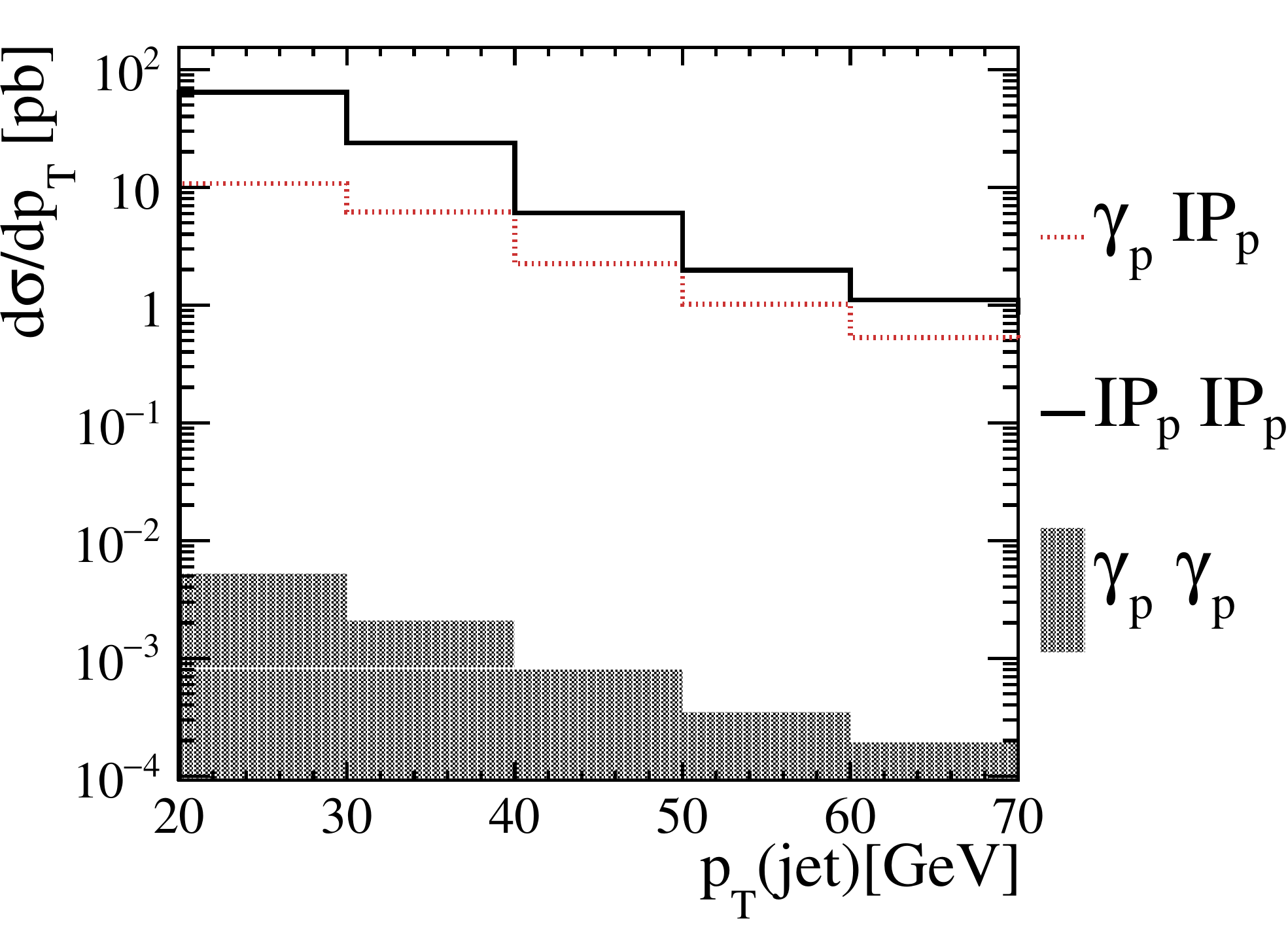}
\end{tabular}
\caption{Differential cross sections as function of $\eta({\rm jet})$ (\textit{left}),   $\Delta\varphi({\rm jet})$ (\textit{center}) and $p_T({\rm jet})$ (\textit{right}) for the dijet production by $\gamma\gamma$, $\gamma\pom$ and $\pom\pom$  interactions in pp collisions. 
}
\label{fig:pp}
\end{center}
\end{figure}

\begin{figure}[t]
\begin{center}
\begin{tabular}{ccc}
\includegraphics[scale=0.3]{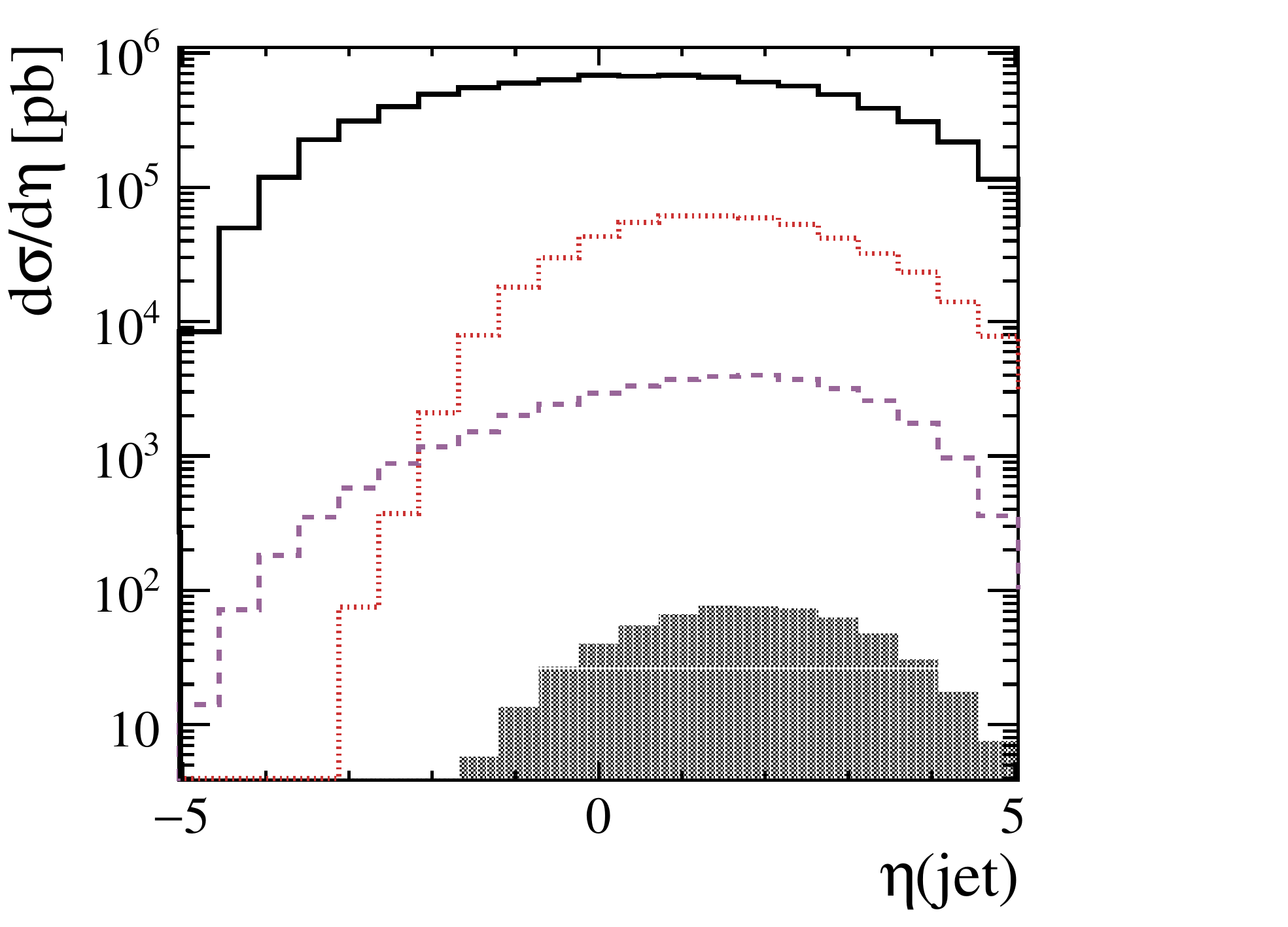}
& \includegraphics[scale=0.3]{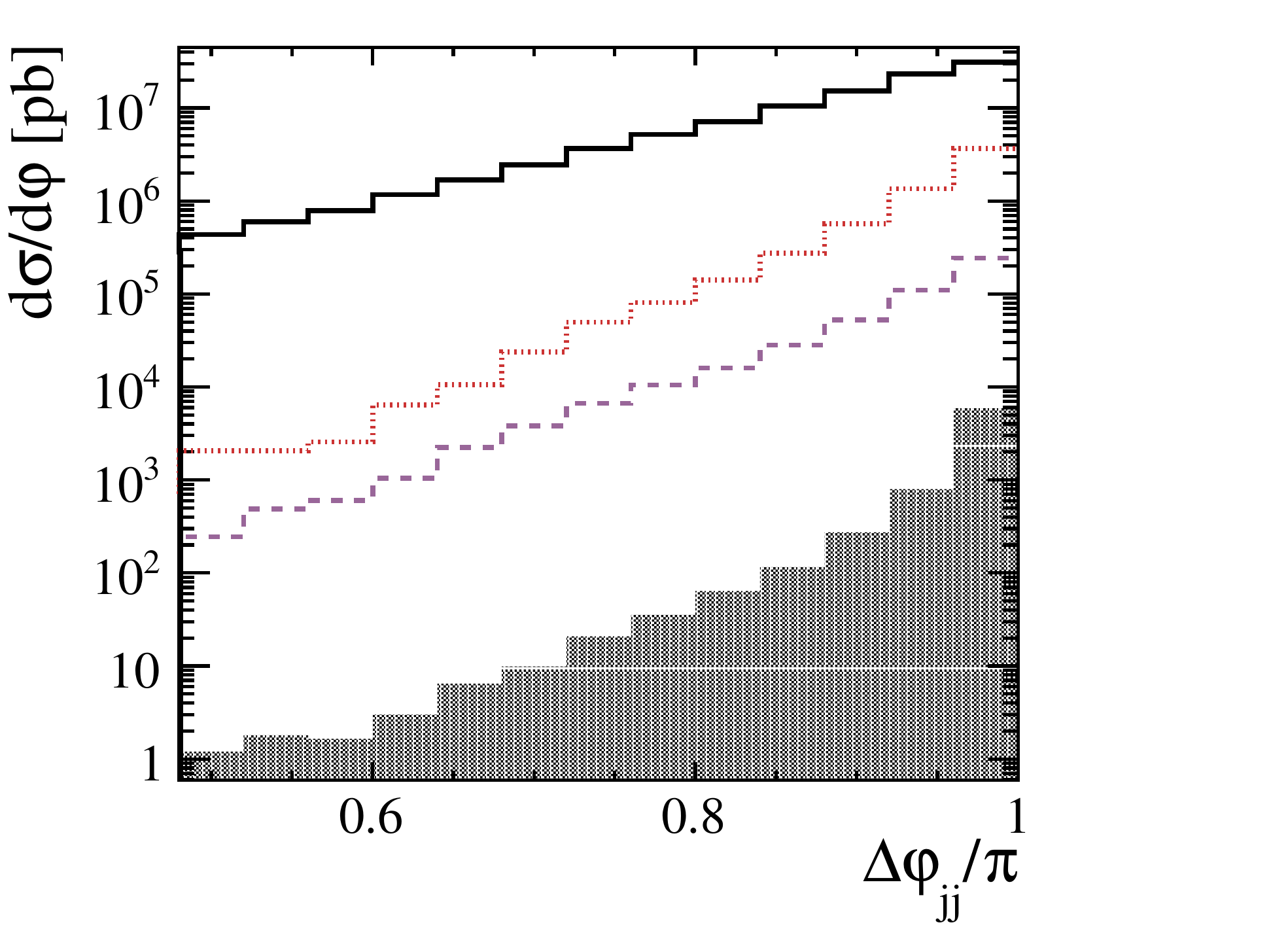} & \includegraphics[scale=0.3]{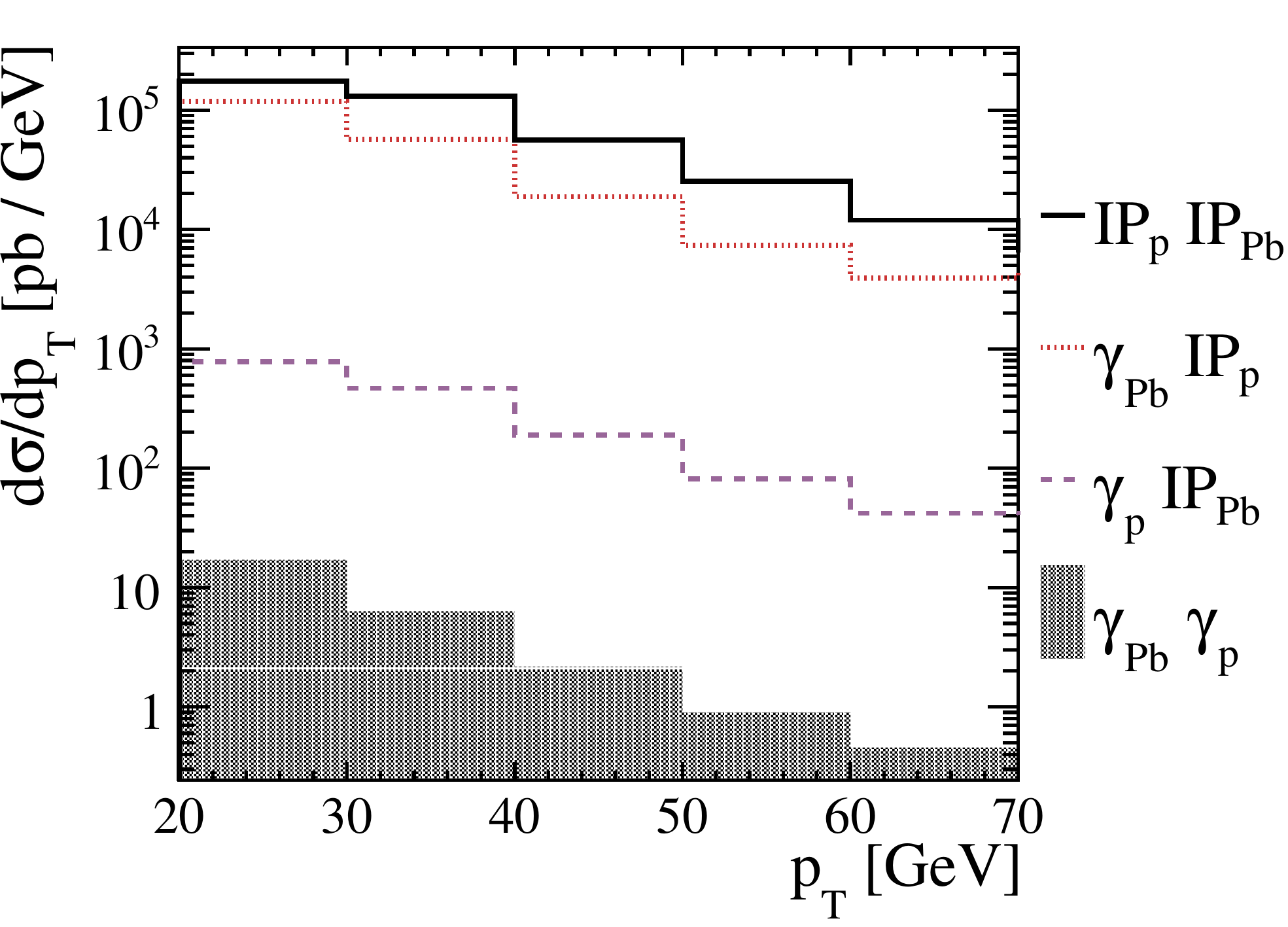}

\end{tabular}
\caption{Differential cross sections as function of $\eta({\rm jet})$ (\textit{left}),   $\Delta\varphi({\rm jet})$ (\textit{center}) and $p_T({\rm jet})$ (\textit{right}) for the dijet production by $\gamma\gamma$, $\gamma\pom$ and $\pom\pom$  interactions in \,{pPb} collisions. The contributions for the $\gamma \pom$ interactions associated to a photon emitted by the nucleus (\,{$\gamma_{Pb} \pom_{p}$}) and by the proton (\,{$\gamma_{p} \pom_{Pb}$}) are presented separately. 
}
\label{fig:pA}
\end{center}
\end{figure}

\begin{figure}[t]
\begin{center}
\begin{tabular}{ccc}
\includegraphics[scale=0.3]{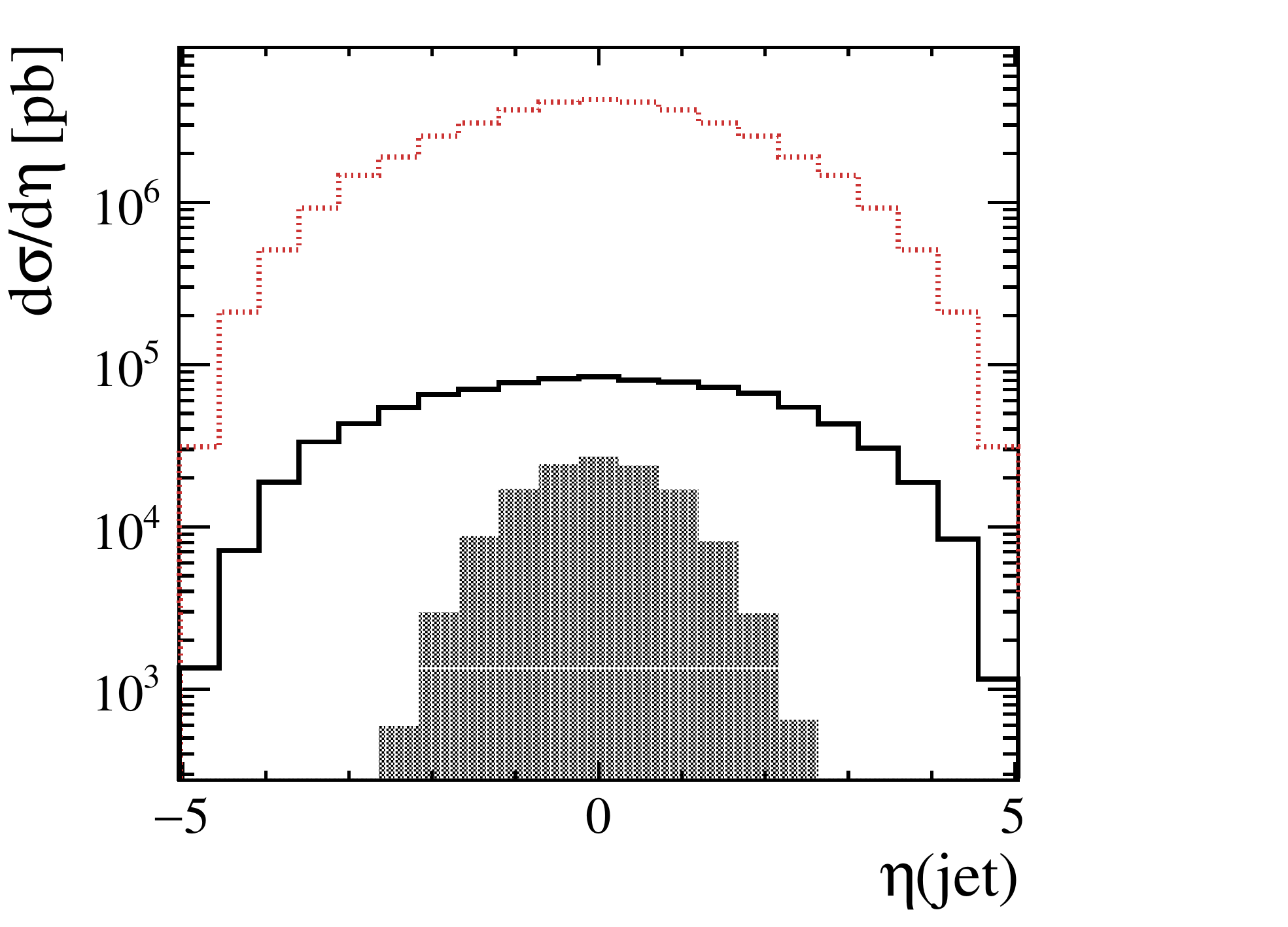}
& \includegraphics[scale=0.3]{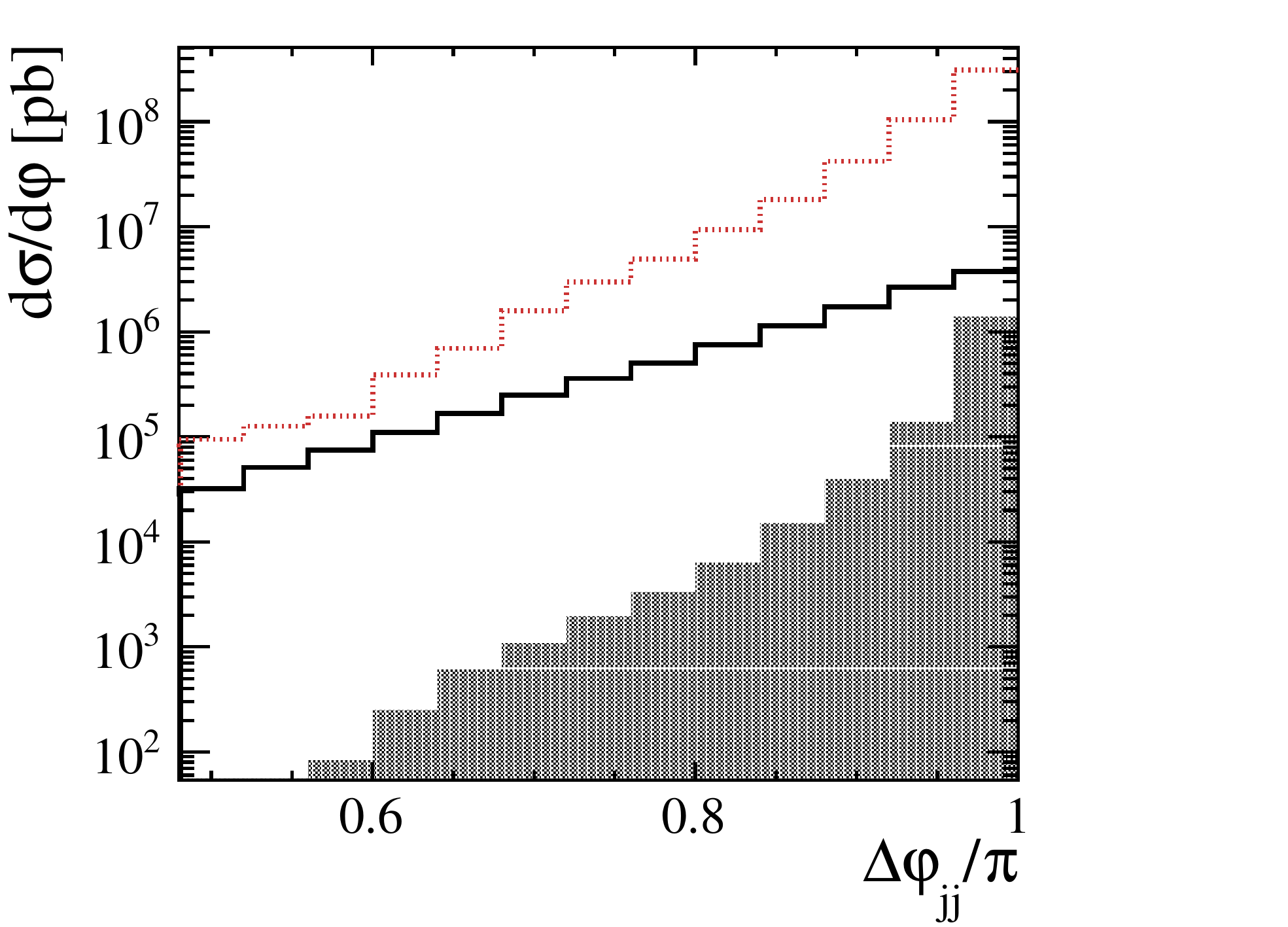}
& \includegraphics[scale=0.3]{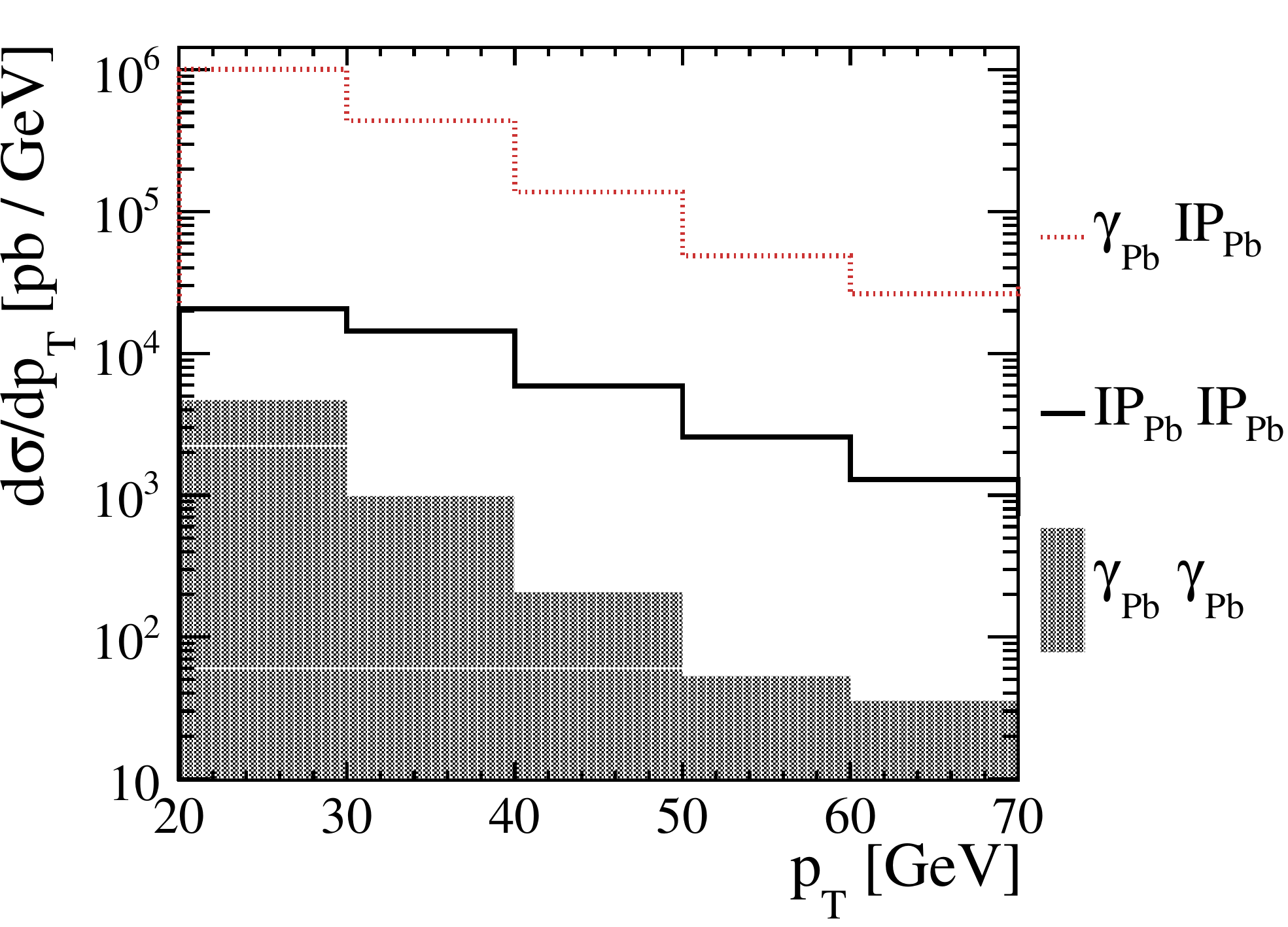}
\end{tabular}
\caption{ Differential cross sections as function of $\eta({\rm jet})$ (\textit{left}),   $\Delta\varphi({\rm jet})$ (\textit{center}) and $p_T({\rm jet})$ (\textit{right}) for the dijet production by $\gamma\gamma$, $\gamma\pom$  and $\pom\pom$  interactions in  \,{PbPb} collisions.  
}
\label{fig:AA}
\end{center}
\end{figure}

\begin{figure}[t]
\begin{center}
\begin{tabular}{ccc}
\includegraphics[scale=0.3]{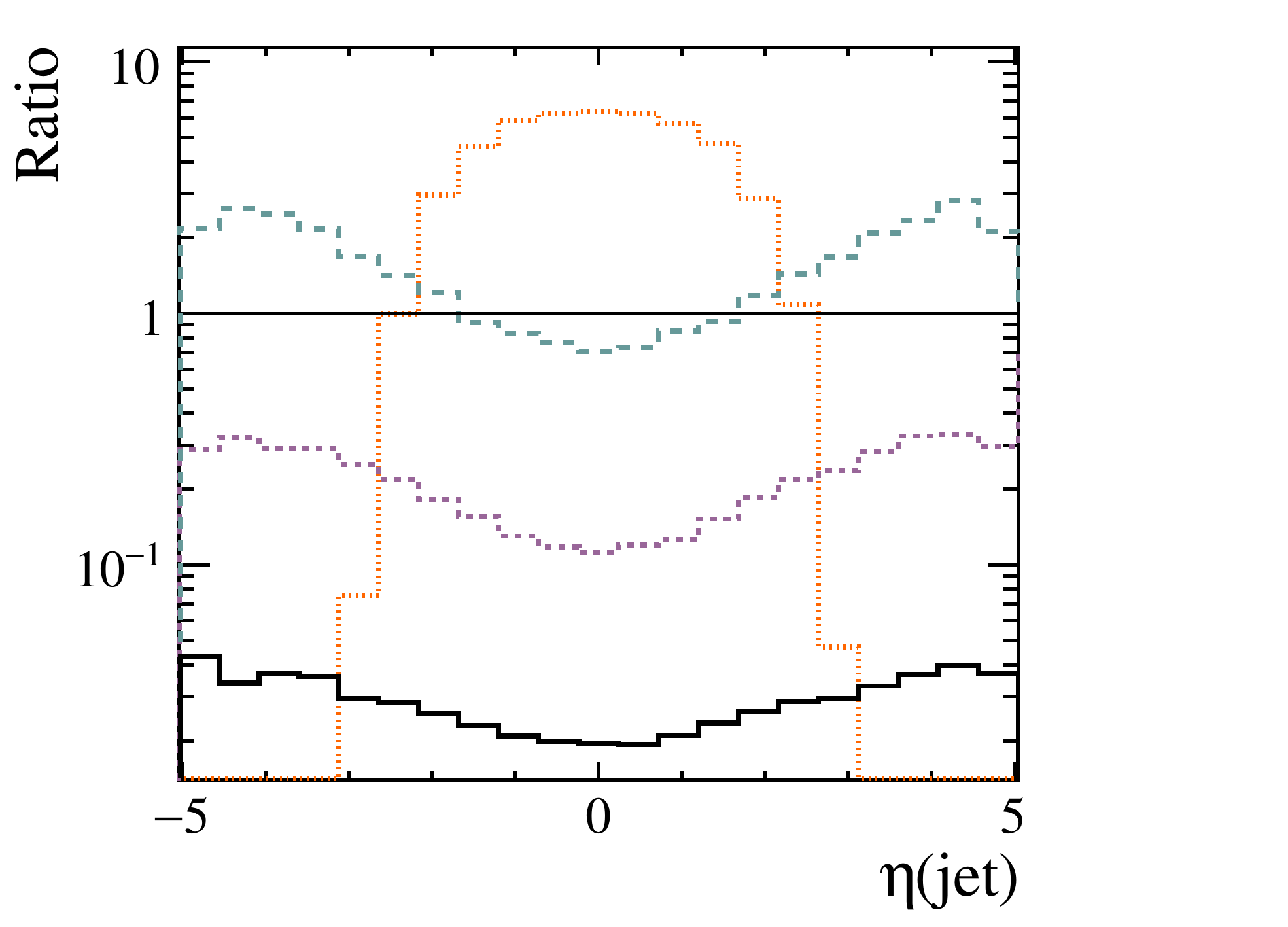}
& \includegraphics[scale=0.3]{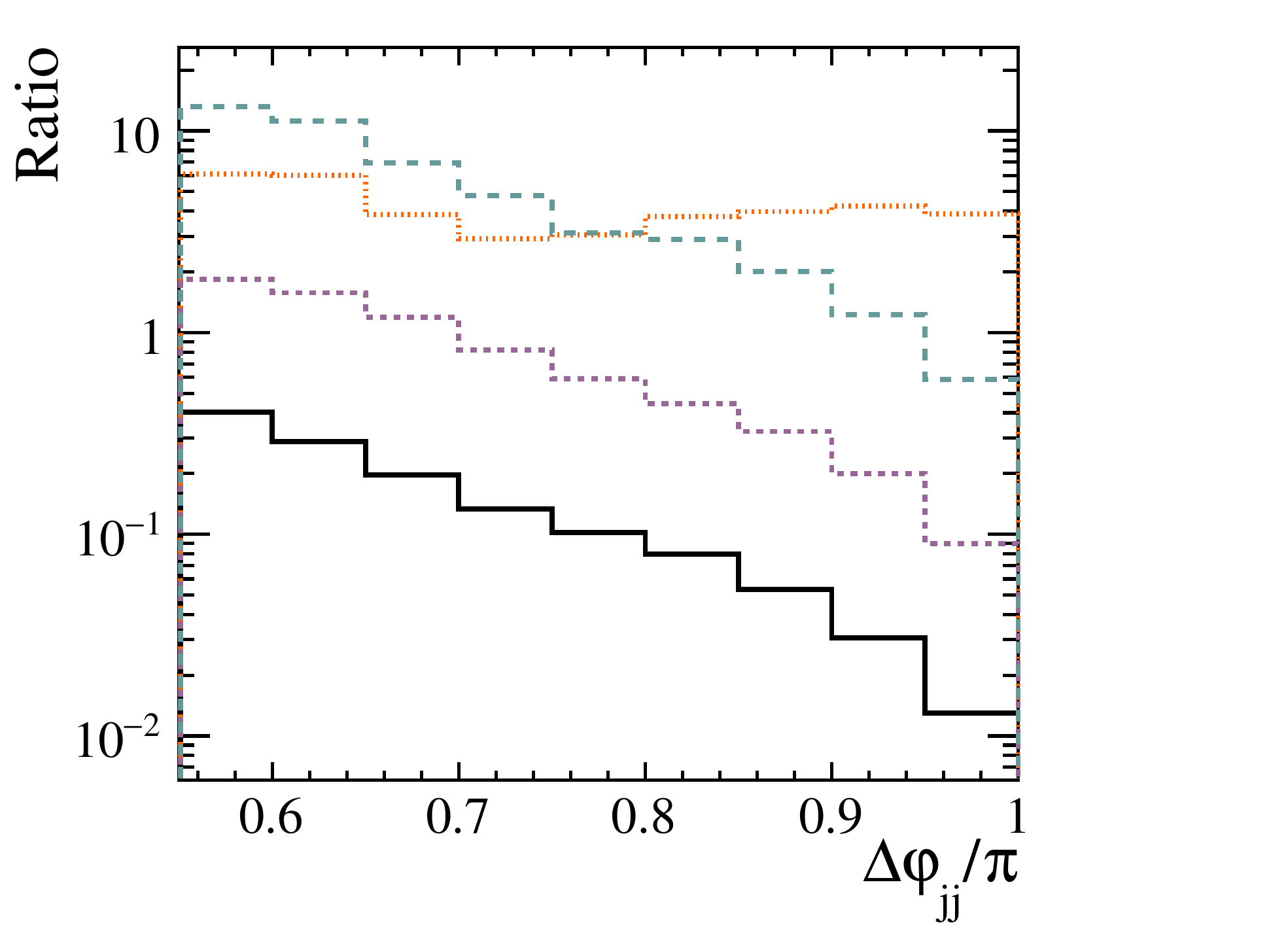}
& \includegraphics[scale=0.3]{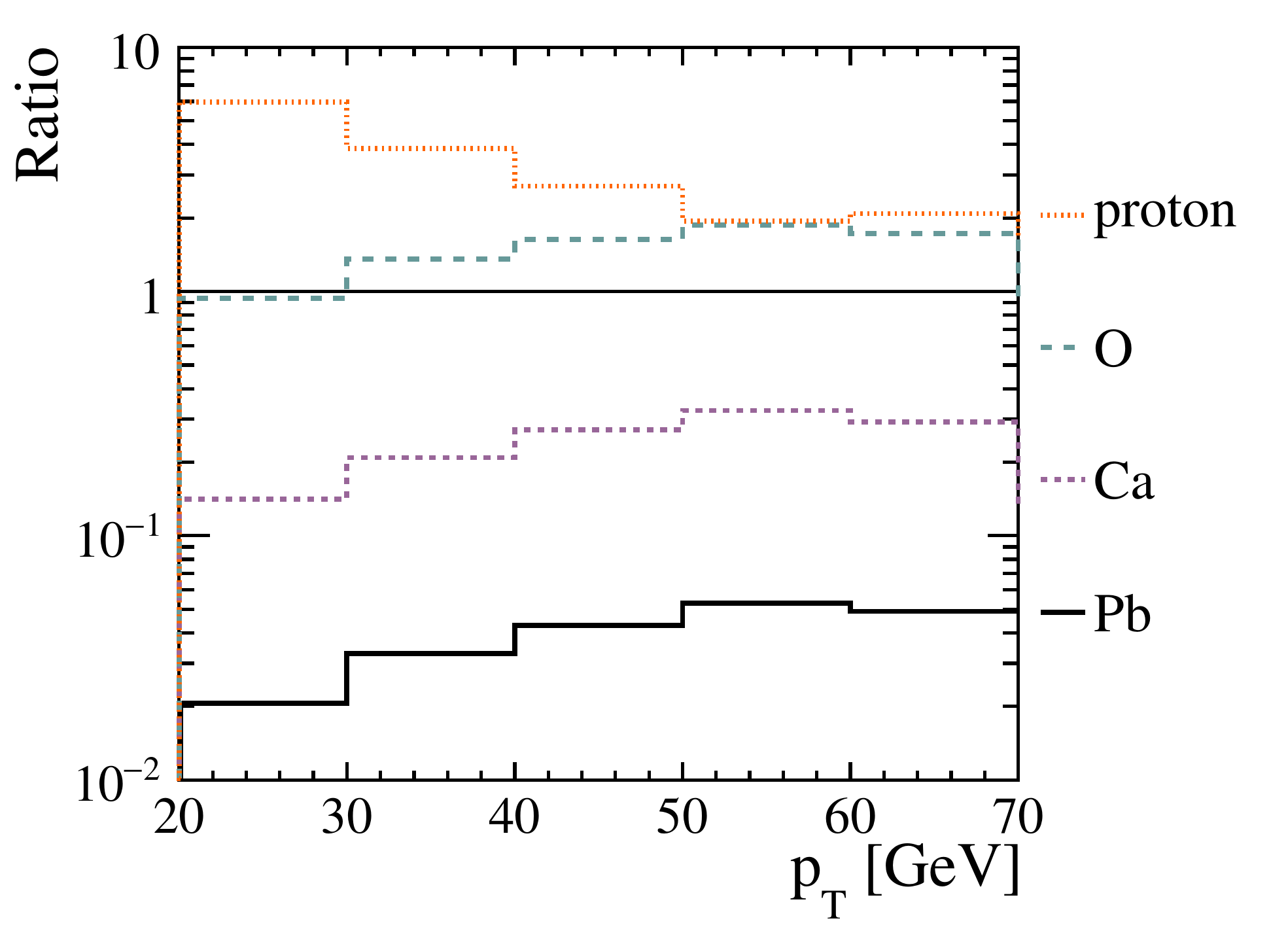} 
\end{tabular}
\caption{ Nuclear contributions on the cross section ratio of the $\pom\pom$ to the $\gamma\pom$ channels as function of $\eta({\rm jet})$ and $\Delta\varphi({\rm jet})$. 
Three nuclear species are used: $^{208}$Pb, $^{40}$Ca, and $^{16}$O. 
}
\label{fig:etaNuclei_ratio}
\end{center}
\end{figure}


\section{Results}
\label{results}

In what follows we present our results for the dijet production  by photon -- photon, photon -- \,{pomeron} and \,{pomeron} -- \,{pomeron}  interactions in \,{pp, pA and AA} collisions at the LHC  energy (For a similar analysis for the heavy quark production Refs. \cite{antoni,nosbottom}). 
As discussed in the Introduction, these processes are characterized by two rapidity gaps and intact hadrons in the final state. The experimental separation of these events using the two rapidity gaps to tag the event  is not an easy task  at the LHC  due to the non - negligible \,{pile-up} present in the normal runs. An alternative  is the detection of the outgoing intact hadrons. Recently, the 
ATLAS, CMS and TOTEM Collaborations have proposed the setup of forward detectors \cite{Albrow:2008pn,ctpps,marek}, which will enhance the kinematic coverage for such investigations. Moreover, the LHCb experiment can study diffractive events by requiring forward regions void of particle production $5.5<|\eta|<8.0$~\cite{Jpsi13tev}.

In our analysis we will assume {pp, pPb and PbPb} collisions at a common center of mass energy ($\sqrt{s} = 5.02$ TeV) in order to estimate their  relative contributions as well as how the different channels of production are modified by increasing the atomic number. Moreover,  we have reconstructed the jet using the anti -- k$_T$ algorithm~\cite{antikt} with distance parameter $R=0.5$ as implemented in the Fastjet software package~\cite{fastjet} and selected jets with $p_T > 20$ GeV and $|\eta|<6.0$. In the case of $\gamma \gamma$ interactions we  include the requirement that the impact parameter of the colliding hadrons should be larger than the sum of its radii. The cross sections for the partonic subprocesses are calculated at leading order in FPMC using HERWIG 6.5. Finally, the subleading contribution for the diffractive interactions associated to the Reggeon exchange will be disregarded in our calculations. As demonstrated e.g. in Refs. \cite{antoni2,marquet3}, such contribution can be similar to the Pomeron one in some regions of the phase space. However, as the description of the Reggeon exchange still is an open question, we postpone the analysis of its impact for the dijet production for a future publication.

 In Fig. \ref{fig:pp} we show the distributions of the pseudo -- rapidity $\eta({\rm jet})$ ({left  panel}) and transverse momentum $p_T({\rm jet})$ (right panel) of the highest-$p_T$ jet, and the azimuthal angular distance between the two highest-$p_T$ jets  $\Delta\varphi_{jj}$. The predictions are presented separately for the dijet production by $\gamma \gamma$, $\gamma \pom$ and $\pom \pom$ interactions in \,{pp} collisions. 
 Initially, lets analyze the $\eta({\rm jet})$ distribution.  We have that the contribution of the $\pom \pom$ process dominates at central pseudo -- rapidities, being a factor $\approx$ 10 ($10^4$) larger than the $\gamma \pom$ ($\gamma \gamma$) one. However, the dijet production by  $\gamma \pom$ interactions implies a broader pseudo -- rapidity distribution. As a consequence, this process  becomes dominant for $|\eta({\rm jet})| \ge 2.8$. In particular, in the kinematical region probed by the LHCb detector, the dijet production will be dominated by $\gamma \pom$ interactions. Consequently, the analysis of this process by the LHCb Collaboration can be an important test for the QCD treatment of the photoproduction of dijets in terms of the Resolved Pomeron model.
On the other hand, the main contribution for the  $p_T({\rm jet})$ distribution comes from the $\pom \pom$ interactions, which is directly associated to the dominance of  central rapidities in the calculation of this distribution. Similarly, the $\pom \pom$ interaction is dominant in the angular distribution of the dijets, with the main contribution being associated to back - to - back configurations.

The results for the dijet production in \,{pPb} collisions are presented in Fig. \ref{fig:pA}. In this case we obtain asymmetric 
$\eta({\rm jet})$ distributions, which is expected since the photon and pomeron fluxes are different for a proton and a nucleus. In order to demonstrate it, we show separately in  Fig.  \ref{fig:pA} the \,{$\gamma_{Pb} \pom_{p}$} and \,{$\gamma_{p} \pom_{Pb}$} contributions, which are associated to a photon emitted by a proton and a nucleus, respectively. As in \,{pp} collisions, the $\pom \pom$ contribution is dominant at central pseudo - rapidities and in the  $p_T$ range considered. Moreover, events are characterized by  back - to - back configurations for the dijets. However, differently form the \,{pp} case, the $\pom \pom$ contribution is larger than the $\gamma \pom$ one in all range of $\eta({\rm jet})$  considered. In particular, for  $\eta({\rm jet}) \le  - 3$, it dominates by a factor $\approx 10^3$, which implies that analysis of the dijet production in this kinematical region can be useful to test the description of $\pom \pom$ interactions in nuclear reactions.  It is important to emphasize that this conclusion is not modified even if our prediction for $\langle |S|^2 \rangle$ is reduced by two orders of magnitude, as predicted in alternative models for the calculation of the gap survival probability in nuclear collisions \cite{radion,miller}.

In Fig. \ref{fig:AA} we present our predictions for the dijet production in \,{PbPb} collisions. In this case we have that the $\gamma \pom$ contribution dominates  in all pseudo - rapidity and \,{transverse} momentum ranges considered. In particular, at central pseudo -- rapidities, we predict that the difference between the predictions is of the order of $10^2$.This result is directly associated to the large suppression of the diffractive interactions in nuclear collisions due to the soft \,{re-scattering} processes that imply the dissociation of the incident nuclei and generate new particles that populate the rapidity gaps in the final state. As a consequence, we have a very small value for the the gap survival probability in \,{PbPb} collisions (See Table \ref{tab:pomnuc}). Although the $\gamma \gamma$ interactions in nuclear collisions are enhanced by a factor $Z^4$ in comparison to \,{pp} case, our results indicate that this channel is only competitive for the dijet production at $\eta({\rm jet}) \approx 0$. In order to estimate the atomic number dependence of the relative contribution between the $\pom \pom$ and $\gamma \pom$ channels for the dijet production, in Fig. \ref{fig:etaNuclei_ratio} we present our predictions for the ratio between $\pom \pom$ and $\gamma \pom$ distributions considering \,{AA} collisions and different values of \,{A}. For comparison, the prediction for \,{pp} collisions also is presented. Our results indicate that the $\pom \pom$ contribution increases at ligher nuclei and become dominant in the dijet production at central rapidities in \,{pp} collisions. On the other hand, the $\gamma \pom$ channel is dominant in \,{CaCa} and \,{PbPb} collisions.    In principle, this conclusion  should not modified by more elaborated models for the calculation of $\langle |S|^2 \rangle$. As discussed before, $\langle |S|^2 \rangle$ in $\gamma \pom$ interactions is expected to be of the order of the unity, while $\pom \pom$ one the alternative models for $\langle |S|^2 \rangle$ in nuclear reactions predict smaller values than that used in our analysis. Therefore, we believe, that the analysis of the dijet production in nuclear collisions with heavy nuclei can be useful to study the photon - \,{pomeron} mechanism at high energies.

Before to summarize our results, some comments are in order. In our calculations we have estimate the dijet production in $\gamma \pom$ interactions considering the leading order subprocesses  present in the HERWIG 6.5 Monte Carlo. The contribution of the next - to - leading order (NLO) corrections for this process is large \cite{kk1}, being approximately a factor 2. The comparison of these predictions with the recent H1 and ZEUS data indicates that the NLO QCD calculations overestimate the data by approximately $40 - 50 \%$ (For a recent review see, e.g. Ref. \cite{alicia}), with the origin of this suppression being a theme of intense debate (See Ref. \cite{kk} for a recent discussion).  As a consequence, we believe that the leading order predictions are a reasonable first approximation for the dijet photoproduction. However, the inclusion of the NLO corrections and a suppression model for $\gamma \pom$ interactions is an important aspect that deserves a more detailed analysis in the future. Another important shortcoming in our study is associated to the fact that we only have considered the direct component of the photon for the dijet photoproduction, where  a point -- like photon interacts with a parton from the Pomeron.  In other words, we have disregarded the resolved component, where the photon behaves as a source of partons, which subsequently interacts with partons from the Pomeron.  
In principle, these two components can be separated by measuring the photon momentum participating in the production of the dijet system, denoted by $x_{\gamma}$. Theoretically, one expect the dominance of the direct (resolved) processes at high (low) values of $x_{\gamma}$. Experimentally, this separation is not so simple due to hadronization and detector resolution and acceptances, but still feasible. 
Therefore,  our calculations for the dijet production in $\gamma \pom$ interactions are realistic for events with large values of $x_{\gamma}$.  
However, as the resolved processes are predicted to be important at small $p_T$ and  large $\eta({\rm jet})$ \cite{klasen_review}, it is possible to analyse the expected impact of the resolved contribution in our main conclusions. In the case of $pp$ collisions (See Fig. \ref{fig:pp}),  the resolved processes should to increase the $\gamma \pom$ prediction for the pseudo - rapidity distribution in the region of large values of $\eta({\rm jet})$, where the $\gamma \pom$ interaction is dominant. Consequently, our conclusion that the production of dijets by $\gamma \pom$ interactions can be studied in $pp$ collisions by the analysis of the large - $\eta({\rm jet})$ region is not expected to be modified by the inclusion of the resolved processes. Similarly, by the analysis from Fig. \ref{fig:AA}, we have that the dominance in $PbPb$ collisions of the   $\gamma \pom$ interactions in the full $\eta({\rm jet})$ range should not be modified. Finally, in the case of $pPb$ collisions, as the  dijet production  by $\pom \pom$ interactions is a factor $\gtrsim 10$ than the direct $\gamma \pom$ prediction (See Fig. \ref{fig:pA}), we also do not expected that this dominance to be modified by the inclusion of the resolved $\gamma \pom$ contribution. Therefore, we believe that our main conclusions must not be strongly modified if this process is included in the analysis. However, we also believe that the resolved contribution for the dijet production is an important aspect that  deserves to be considered and we plan to include this contribution in the FPMC generator.

\section{Summary}
\label{conc}

As a summary, in this paper we have presented a detailed analysis  for the dijet production  in \,{pp/pA/AA} collisions at the LHC. In particular, the comparison between the predictions for the dijet production by  photon -- photon, photon -- pomeron and pomeron -- pomeron  interactions was presented  considering a common framework implemented in the Forward Physics Monte Carlo. We have generalized this Monte Carlo for nuclear reactions and performed a detailed comparison between the 
 $\pom \pom$,  $\gamma \pom$ and $\gamma \gamma$ predictions for the dijet production in \,{pp/pPb/PbPb} collisions at $\sqrt{s} = 5.02$ TeV. For the pomeron - induced processes in \,{pp} collisions, we have considered the  framework of the Resolved Pomeron model corrected for absorption effects, as used in the estimation of several other diffractive processes.  In the case of nuclear collisions, we have generalized this model, following Refs. \cite{berndt,vadim}. Moreover, the absorption effects also have been included in our estimates for the dijet production by   $\pom \pom$ interactions in nuclear collisions. Our results indicate that in \,{pp} collisions the $\pom \pom$ channel is dominant  at central rapidities, being suppressed at forward rapidities. In particular, in the kinematical range probed by the LHCb detector, we predict that the main contribution for the dijet production comes from $\gamma \pom$ interactions. In the case of \,{pPb} collisions, the $\pom \pom$ interactions are dominant. In contrast, our results indicated that in \,{AA} collisions with heavy nuclei, the dijet production by $\gamma \pom$ interactions is dominant, which indicates that this process can be used to test the Pomeron Resolved Model and its generalization for nuclei.
 Finally, our results indicate that the experimental analysis of the dijet production  would help to constrain the underlying model for the \,{pomeron} and the absorption corrections, which are important open questions in Particle Physics.

\begin{acknowledgments}
Useful discussions with Marek Tasevsky are gratefully acknowledged. This research was supported by CNPq, CAPES, FAPERJ and FAPERGS, Brazil. 
\end{acknowledgments}

\appendix

\section{{Modelling $\langle S^2 \rangle$  in the impact parameter space}}

In order to estimate the survival gap probability in the impact parameter space, we will  consider an approach similar to that proposed in Refs. \cite{RAU:1990,berndt}. In this Appendix we present the basic aspects of this approach and postpone for a future publication a detailed discussion of the assumptions and uncertainties present in our calculations. Initially,  lets to express  Eq.(\ref{pompom}) in terms of the Pomeron-Pomeron cross section
\begin{eqnarray}
\sigma(h_{\rm A} h_{\rm B} \rightarrow h_{\rm A} \otimes X j_1 j_2 Y \otimes h_{\rm B}) = \int dx_{\rm A} \int dx_{\rm B} \, ~\sigma(\pom_{\rm A}\pom_{\rm B}\rightarrow X j_1 j_2 Y ) ~,
\label{pompom:2}
\end{eqnarray}
where 
\begin{eqnarray}
\sigma(\pom_{\rm A}\pom_{\rm B}\rightarrow X j_1 j_2 Y ) = \int_{x_{\rm A}}^1 \frac{dx_{\pom_{\rm A}}}{x_{\pom_{\rm A}}} \int_{x_{\rm B}}^1 \frac{dx_{\pom_{\rm B}}}{x_{\pom_{\rm B}}} g_{\pom/{\rm A}}\left(\frac{x_{\rm A}}{x_{\pom_{\rm A}}}, \mu^2\right)g_{\pom/{\rm B}}\left(\frac{x_{\rm B}}{x_{\pom_{\rm B}}}, \mu^2\right)f_{\pom/{\rm A}}(x_{\pom_{\rm A}})~f_{\pom/{\rm B}}(x_{\pom_{\rm B}})\cdot \hat{\sigma}(g g \rightarrow j_1 j_2) ~.
\label{pompom:3}
\end{eqnarray}
Taking into account the transferred momentum $q^2$  dependence of the partons emitted by the Pomeron, the above equation can be written in terms of the Pomeron flux in the momentum space ${\bar{f}}_{\pom/{\rm i}}(x_{\pom},q^2)$ as follows
\begin{eqnarray}
&&\sigma(\pom_{A}\pom_{B}\rightarrow X j_1 j_2 Y ) = \int d^2b \int_{x_{\rm A}}^1 \frac{dx_{\pom_{\rm A}}}{x_{\pom_{\rm A}}} \int_{x_{\rm B}}^1 \frac{dx_{\pom_{\rm B}}}{x_{\pom_{\rm B}}} g_{\pom/{\rm A}}\left(\frac{x_{\rm A}}{x_{\pom_{\rm A}}}, \mu^2\right)g_{\pom/{\rm B}}\left(\frac{x_{\rm B}}{x_{\pom_{\rm B}}}, \mu^2\right)\times\nonumber \\
&& \int \frac{d^2q}{(2\pi)^2}e^{iq.b}
\bar{f}_{\pom/{\rm A}}(x_{\pom_{\rm A}},q^2)~\bar{f}_{\pom/{\rm B}}(x_{\pom_{\rm B}},q^2)\cdot \hat{\sigma}(g g \rightarrow j_1 j_2) ~.
\label{pompom:4}
\end{eqnarray} 
with 
\begin{eqnarray}
\bar{f}_{\pom/{\rm A}}(x_{\pom},q^2)=R_g {\rm A}^2\int \frac{d^2k}{(2\pi)^2} F_{\rm A}(k^2)F_{\rm A}((k-q)^2)\Delta_{\pom}(x_{\pom},k^2)\Delta_{\pom}(x_{\pom},(k-q)^2)~,
\label{q_dependent_pdf_pomA}
\end{eqnarray}
and
\begin{eqnarray}
\bar{f}_{\pom/{\rm B}}(x_{\pom},q^2)=R_g {\rm B}^2\int \frac{d^2k}{(2\pi)^2} F_{\rm B}(k^2)F_{\rm B}((k+q)^2)\Delta_{\pom}(x_{\pom},k^2)\Delta_{\pom}(x_{\pom},(k+q)^2)~.
\label{q_dependent_pdf_pomB}
\end{eqnarray}
 The functions $F_i$ and $\Delta_{\pom}$ characterize the hadronic form factors  and Pomeron propagators, respectively. 
In what follows we  will assume that 
\begin{eqnarray}
F_{\rm i}(k^2)= A_{\pom}^{1/2}e^{R_{\rm i}^2 k^2/6}~. 
\end{eqnarray}
and 
\begin{eqnarray}
\Delta_{\pom}(x_{\pom},k^2)=x_{\pom}^{-\frac{1}{2}-\epsilon}e^{-\alpha'\log x_{\pom}~ k^2 }~,
\end{eqnarray}
with the parameters being those obtained by the HERA H1 experiment \cite{H1diff}.
As both the form-factor and the propagator are Gaussians in $k^2$,  the integrals over the Pomeron momentum $k$ can be performed. Using the above forms in Eqs.(\ref{q_dependent_pdf_pomA},\ref{q_dependent_pdf_pomB}) we write
\begin{eqnarray}
\bar{f}_{\pom/{\rm A}}(x_{\pom},q^2)= R_g {\rm A}^2\times A_{\pom}~x_{\pom}^{-1-2\epsilon}\int \frac{d^2k}{(2\pi)^2}~e^{\bar{R}_{A}^2 k^2}e^{\bar{R}_{A}^2 (k-q)^2}  ~,
\label{q_dependent_pdf_pom_1}
\end{eqnarray}
and
\begin{eqnarray}
\bar{f}_{\pom/{\rm B}}(x_{\pom},q^2)= R_g {\rm B}^2\times A_{\pom}~x_{\pom}^{-1-2\epsilon}\int \frac{d^2k}{(2\pi)^2}~e^{\bar{R}_{B}^2 k^2}e^{\bar{R}_{B}^2 (k+q)^2}  ~,
\label{q_dependent_pdf_pom_2}
\end{eqnarray}
with $
\bar{R}_{\rm A}^2=\frac{R_{\rm A}^2}{6} -\alpha'\log x_{\pom}~,
$ and
$\bar{R}_{\rm B}^2=\frac{R_{\rm B}^2}{6} -\alpha'\log x_{\pom}~.
$
Using Eqs. (\ref{q_dependent_pdf_pom_1}), (\ref{q_dependent_pdf_pom_2}) and (\ref{pompom:4}) in Eq. (\ref{pompom:3}) and performing the integrations over $k_{\rm A}$,$k_{\rm B}$ and $q$ variables we obtain
 \begin{eqnarray}
\frac{d\sigma}{d^2b}(\pom_{\rm A}\pom_{\rm B}\rightarrow X j_1 j_2 Y ,~b)=\frac{1}{2\pi}\tilde{Q}^2e^{-\tilde{Q}^2b^2/2}\sigma(\pom_{\rm A}\pom_{\rm B}\rightarrow X j_1 j_2 Y)~,
 \label{pomeron_pomeron sigma_1}
\end{eqnarray}
where
\begin{eqnarray}
\tilde{Q}^2 \approx \frac{6}{{R}_{\rm A}^2+{R}_{\rm B}^2}~.
\end{eqnarray}
In order to calculate the integrated cross section taking into account the absorptive  effects we multiply the above by the probability of not having strong interactions $e^{-\Omega(s,b)}$, where $\Omega(s,b)$ is the nuclear/proton opacity. 
In the $pp$ case, we assume that the proton elastic profile can be describe by a Gaussian form, which implies that the opacity in proton-proton collisions is given by
\begin{equation}
\Omega_{pp}(s,b)=\frac{\sigma_{\rm tot}(s)}{4\pi}\frac{2}{B_{\rm soft}(s)}e^{-b^2/2B_{\rm soft}(s)}~,
\label{proton_opacity}
\end{equation}
where $\sigma_{\rm tot}$ is the total pp cross section and $B_{\rm soft}$ is the elastic scattering effective slope. We take these parametrizations from \cite{KFK:2014}. 
In order to derive a similar expression for the nuclear case, which is simple and can be used in analytical calculations, we have adjusted the Wood - Saxon distribution for the nuclei by a Gaussian one $\propto e^{-Q_0^2b^2/4}$, with $Q_0$ being an effective  parameter fitted to each nuclei. A similar procedure was proposed in Ref. \cite{dress}.   As a consequence, we obtain for $AA$ collisions that  
\begin{equation}
\Omega_{\rm AA}(s,b)={\rm A}^2\frac{\sigma_{\rm tot}(s)}{4\pi}Q_0^2e^{-Q_0^2b^2/4}~,
\label{nuclear_opacity}
\end{equation}
where A  is the atomic number and $Q_0$ is obtained from the nuclear form factor \cite{dz}. On the other hand, for  pA collisions we consider that the opacity can be expressed  by 
\begin{equation}
\Omega_{pA}(s,b)={\rm A}\frac{\sigma_{\rm tot}(s)}{4\pi B_{\rm eff}}e^{-b^2/4B_{\rm eff}}~,
\label{Ap_opacity}
\end{equation}
where $B_{\rm eff}=\frac{1}{2Q_0^2}+\frac{B_{\rm soft}}{4}$.
 
\begin{figure}[t]
\begin{center}
\begin{tabular}{ccc}
\includegraphics[scale=0.3]{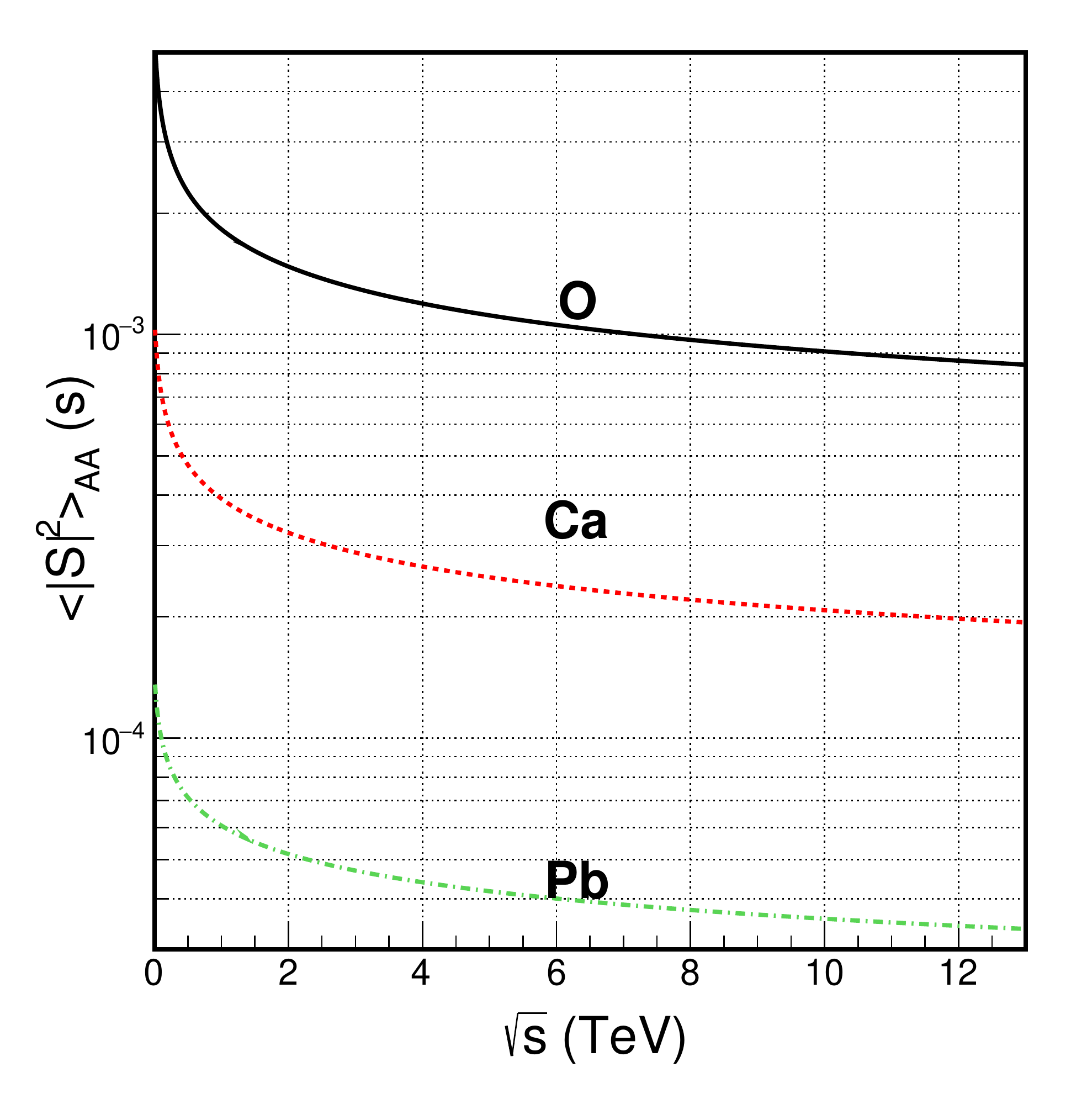}
\includegraphics[scale=0.3]{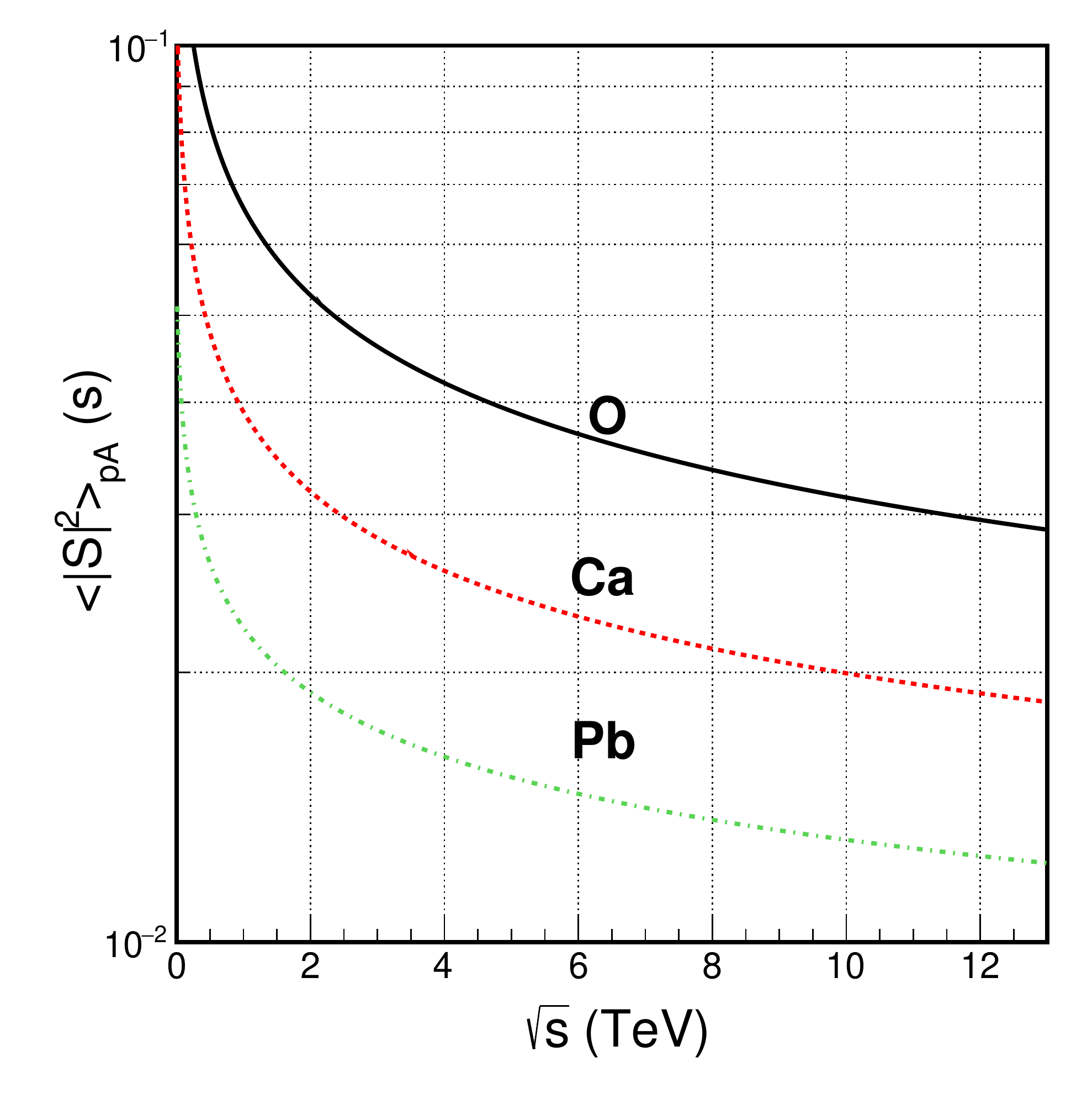}
\end{tabular}
\caption{Energy dependence of the survival gap factors $\langle |S_{\mathbb{P}_{\rm A}\mathbb{P}_{\rm A}}(s)|^2 \rangle$ and $\langle |S_{\mathbb{P}_{\rm p}\mathbb{P}_{\rm A}}(s)|^2 \rangle$ for AA (left panel)  and pA (right panel) collisions considering different nuclei. }
\label{survival_factor}
\end{center}
\end{figure}

Using the above opacities, it is possible to calculate the  integrated cross section for a  general $h_a h_b$ collision ($a = p$ or $A$), which will be given by
\begin{eqnarray}
\sigma(\pom_{\rm a}\pom_{\rm b}\rightarrow X j_1 j_2 Y )&=&\int_{0}^{\infty}d^2b~\frac{d\sigma}{d^2b}(\pom_{\rm a}\pom_{\rm b}\rightarrow X j_1 j_2 Y ,~b)e^{-\Omega(s,b)}= \langle |S_{\pom_{\rm a}\pom_{\rm b}}|^2 \rangle \times\sigma(\pom_{\rm a}\pom_{\rm b}\rightarrow X j_1 j_2 Y )|_{\Omega = 0} ~. \nonumber \\
 \label{pomeron_pomeron_sigma_2}
\end{eqnarray}
For AA collisions, the suppression factor can be expressed by
\begin{eqnarray}
\langle |S_{\pom_{  A}\pom_{  A}}(s)|^2 \rangle =\frac{\tilde{Q}^2}{2}~\int_{0}^{\infty}db^2~e^{-\tilde{Q}^2b^2/2}\exp\Big(-{  A}^2\frac{\sigma_{  tot}(s)}{4\pi}Q_0^2e^{-Q_0^2b^2/4}\Big) ~,
 \label{suppresion_pom-pom}
\end{eqnarray}
with $\tilde{Q}^2 \approx Q_0^2$. As the above integral is of the type
\begin{eqnarray}
I=\int_0^{\infty} dx^2~ e^{-\alpha x^2} e^{-\lambda e^{-\beta x^2}}=-\frac{1}{\alpha}\int_{1}^{0}dy~\exp(-\lambda~y^{\beta/\alpha})=\frac{\lambda^{-\alpha/\beta}}{\beta}~\gamma\Big(\frac{\alpha}{\beta},\lambda\Big)~,
\end{eqnarray}
where $\gamma$ is the incomplete gamma function, $\alpha=\tilde{Q}^2/2$, $\lambda=A^2\frac{\sigma_{  tot}}{4\pi}Q_0^2$ and $\beta=Q_0^2/4$, $\langle |S_{\pom_{  A}\pom_{  A}}(s)|^2 \rangle$ can be  written as follows
\begin{eqnarray}
\langle |S_{\pom_{  A}\pom_{  A}}(s)|^2 \rangle=\frac{2\tilde{Q}^2}{Q_0^2}\Big({  A}^2\frac{\sigma_{  tot}}{4\pi}Q_0^2\Big)^{-2\tilde{Q}^2/Q_0^2}\gamma\Big(\frac{2\tilde{Q}^2}{Q_0^2}, {  A}^2\frac{\sigma_{  tot}}{4\pi}Q_0^2\Big) ~.
 \label{suppresion_pom-pom-AA}
\end{eqnarray}
Similarly, we can obtain the suppression factor for  pA collisions, which is given by
\begin{eqnarray}
\langle |S_{\pom_{p}\pom_{  A}}(s)|^2 \rangle =2\tilde{ Q}^{2}_{  eff} B_{  eff}\Big({  A}\frac{\sigma_{  tot}}{4\pi B_{  eff}}\Big)^{-2\tilde{Q}^2 B_{  eff}}\gamma\Big(2\tilde{Q}^2 B_{  eff}, {  A}\frac{\sigma_{  tot}}{4\pi B_{  eff}}\Big) ~,
 \label{suppresion_pom-pom-pA}
\end{eqnarray}
where $\tilde{ Q}^{2}_{eff}=6/(R_A^2+3B_{\pom})$. Finally, the expression for pp collisions is given by
\begin{eqnarray}
\langle |S_{\pom_{p}\pom_{p}}(s)|^2 \rangle = \frac{B_{  soft}}{B_{\pom}}\Big(  \frac{\sigma_{  tot}}{2\pi B_{  soft}}\Big)^{-B_{  soft}/B_{\pom}}\gamma\Big(\frac{B_{  soft}}{B_{\pom}}, \frac{\sigma_{  tot}}{2\pi B_{  soft}}\Big) ~,
 \label{suppresion_pom-pom-pp}
\end{eqnarray}
which is similar to the expression derived in Ref. \cite{GLevin:2006} using a distinct approach. Using the above expressions we have derived the values for the survival gap probabilities for Pomeron -- Pomeron interactions in $PbPb$ and $pPb$ collisions presented in Table \ref{tab:pomnuc}. In Fig. \ref{survival_factor} we present our predictions for its energy dependence considering $AA$ and $pA$ collisions and different nuclei.   It is important to emphasize that our prediction for $pp$ collisions at the LHC energy is $\approx 0.02$, as used in several phenomenological analysis in the literature \cite{nosbottom,MMM1,marquet3,antoni,antoni2,cristiano,cristiano2}. 

\end{document}